\documentclass[useAMS,usenatbib,usegraphicx]{mn2e}

\usepackage{upgreek}
\usepackage{amsmath}
\usepackage{txfonts}
\usepackage{booktabs}
\usepackage{color}
\usepackage{array}
\usepackage{graphicx}
\usepackage{epstopdf}
\usepackage{color}
\usepackage{bm}
\usepackage{cmap}
\usepackage{tabularx}
\usepackage{multicol}
\usepackage{natbib}
\usepackage{float}
\usepackage{multirow}

\DeclareMathOperator\erf{erf}
\begin{document}
\title[Statistical theory of neutron star cooling]{
Statistical theory of thermal evolution of neutron stars -- II.\\
Limitations on direct Urca threshold}

\author[M. V. Beznogov and D. G. Yakovlev]{
M. V. Beznogov$^{1}$\thanks{E-mail: mikavb89@gmail.com},
D. G. Yakovlev$^{2}$
\\
$^{1}$St.~Petersburg Academic University, 8/3 Khlopina st.,
St.~Petersburg 194021, Russia \\
$^{2}$Ioffe Physical Technical Institute, 26 Politekhnicheskaya st.,
St.~Petersburg 194021, Russia}

\date{Accepted . Received ; in original form}
\pagerange{\pageref{firstpage}--\pageref{lastpage}} \pubyear{2015}
\maketitle \label{firstpage}

%%%%%%%%%%%%%%%%%%%%%%%%%%%%%%%%%%%%%%%%%%%%%%
\begin{abstract}
We apply our recently suggested statistical approach to thermal evolution of isolated neutron stars and accreting quasistationary neutron stars in X-ray transients for constraining the position and relative broadening $\alpha$ of the direct Urca threshold of powerful neutrino emission in neutron star cores. We show that most likely explanation of observations corresponds to  $\alpha \approx 0.08-0.10$ and to the neutron star mass, at which the direct Urca process is open, $M_{\rm D}\approx 1.6-1.8\,{\rm M}\odot$.
\end{abstract}


%%%%%%%%%%%%%%%%%%%%%%%%%%%%%%%%%%%%%%%%%%%%%
\section{Introduction}
\label{sec:intro}

It is well known that theories of thermal evolution of neutron stars have potential to explore properties of superdense matter in neutron star cores (\citealt{PLPS04,Page_etal09,YP04}). Recently we have suggested a statistical approach  (\citealt{BY15}; hereafter Paper I) to study thermal evolution of middle aged ($t \sim 10^2-10^6$ yr) isolated neutron stars (INSs) and old ($t \gtrsim 10^8$ yr) transiently accreting quasistationary neutron stars in low-mass X-ray binaries (in X-ray transients; XRTs). INSs cool down loosing their thermal energy; neutron stars in XRTs are warmed up by deep crustal heating (\citealt{HZ90,HZ08,BBR98}) during accretion episodes. Thermal evolution of INSs and neutron stars in XRTs allows one to test the same physics of superdense matter in neutron star cores (e.g., \citealt{YH03,YLH03}). Traditional theories of thermal evolution of these objects are based on calculations of cooling curves for INSs (e.g., $T_\mathrm{s}^\infty(t)$), and heating curves for accreting neutron stars (e.g., $L_\gamma^\infty(\langle\dot{M} \rangle ))$. Here, $T_\mathrm{s}^\infty$ is the redshifted effective surface temperature of the star, $t$ is the stellar age, $L_\gamma^\infty$ is the redshifted photon surface luminosity of a neutron star in quiescent states of XRT, and $\langle \dot{M} \rangle$ is the mass accretion rate in XRT averaged over neutron star cooling time-scales (about 1 kyr or more).

The statistical approach of Paper I replaces theoretical cooling/heating curves by probabilities to find neutron stars in certain regions of the $T_\mathrm{s}^\infty-t$ or $L_\gamma^\infty- \langle \dot{M} \rangle$ planes. These probabilities are obtained by averaging individual cooling/heating curves over mass distributions of isolated or accreting neutron stars and over distributions of other neutron star parameters, particularly, of mass of light elements in the heat blanketing envelopes. Light elements increase heat transparency of the heat blanketing envelopes in comparison with standard iron envelopes (e.g., \citealt{GPE83,Potekhin_etal97}) and affect thus the thermal evolution of neutron stars. Taken an increasing statistics of the sources, calculated probabilities simplify comparison of the theory with observations. Moreover, these probabilities depend not only on the properties of superdense matter in neutron stars but also on distributions of neutron star parameters (first of all, on the mass distributions) which enables one to study these distributions together with the properties of superdense matter (Paper I).

Full implementation of the statistical approach is a complicated task. In Paper I the idea was illustrated using neutron star models with nucleon cores having one equation of state (EOS) of superdense matter. This EOS was constructed by \citet{KKPY14} and abbreviated as HHJ because it belongs to the family of EOSs suggested by \citet{HHJ99}. A preliminary analysis of Paper I indicated that observational data on isolated and accreting neutron stars could be reconciled if powerful direct Urca process of neutrino emission \citep{LPPH91} operated in the cores of massive stars and the threshold for the onset of the direct Urca process with growing density $\rho$ were broadened (e.g. \citealt{YKGH01}). Otherwise the theory cannot explain a number of observed sources. Typical masses of INSs need to be lower than the mass $M_{\rm D}$ at which the direct Urca process is on, but accreting neutron stars should be overall more massive to explain at least one source, SAX J1808.4--3658 \citep{Campana_etal05,GC06,Heinke09b}, where the direct Urca process seems to operate.

This paper extends the analysis of Paper I in two ways. First, we consider two EOSs of superdense matter instead of one which illustrates the effects of EOSs. Secondly, we take into account that one should distinguish a formal threshold of the direct Urca process (at a density $\rho=\rho_{\rm D0}$ at which the process becomes allowed by momentum conservation of reacting particles for a particular EOS; \citealt{LPPH91}) and the actual threshold $\rho_{\rm D}$ which can be shifted with respect to $\rho_{\rm D0}$. The shift can be produced, for instance, by superfluidity of neutron star matter; superfluidity can suppress the direct Urca process and increase $\rho_{\rm D}$ (e.g., \citealt{YKGH01,YP04}). In addition, the shift can be produced by strong magnetic fields in neutron stars cores \citep{BY99,YKGH01}; the fields modify momentum conservation rules and can decrease $\rho_{\rm D}$ with respect to $\rho_{\rm D0}$. It is also possible that a threshold for enhanced neutrino emission can be broadened and shifted by nuclear physics effects (as
discussed, for instance, by \citealt{Schaab_etal97,Blaschke_etal04} and in references therein). Let us stress that it is the actual threshold $\rho_{\rm D}$ which regulates thermal evolution of neutron stars; this threshold is basically unknown even if we knew the EOS. Therefore, we will treat $\rho_{\rm D}$ as a free parameter and try to constrain its values. Note that the same effects which shift $\rho_{\rm D}$ can broaden the direct Urca threshold. Thus, although we consider threshold broadening and shifting as independent phenomena, it may be actually not so. In addition, it is important to stress that large shifts and strong broadening seem unlikely from theoretical points of view.

As in Paper I, we will treat the broadening of the direct Urca threshold on phenomenological level by introducing the same model broadening function characterized by the relative broadening width
$\alpha$. We will not use any specific physical  broadening model (superfluidity, magnetic fields, nuclear physics effects) but postpone such studies till future, more detailed consideration. Our present analysis will be qualitative but hopefully it will give general understanding of the problem. The observational basis will be the same as in Paper I (where detailed list of references is given). For convenience of the reader Table \ref{tab:observa} lists observational sources plotted in the figures below.

%%%%%%%%%%%%%%%%%%%%%%%%%%%
\renewcommand{\arraystretch}{1.10}
\begin{table}
\centering 
\caption{Middle-aged cooling INSs (left) and accreting neutron stars in XRTs (right) whose thermal surface emission has been detected or constrained; see Paper I for details.}
    \begin{tabular}{c l | c l}
    %\toprule
   \hline
    Num.  &  INS  & Num. & XRT \\ 
    \hline
    %\midrule
    1  &  PSR J1119--6127 &    1     &  Aql X-1
 \\
    2  &  RX J0822--4300  &  2     &  4U 1608--522
\\
       & (in Pup A)  & 3     &  MXB 1659--29
\\
    3  & PSR J1357--6429  &     4     &  NGC 6440 X-1
\\
    4  & PSR B0833--45 &  5   &  RX J1709--2639
\\
       & (Vela)  &   6   &  IGR 00291+5934
\\
    5  &  PSR B1706--44  &  7     &  Cen X-4
\\
    6  & PSR J0538+2817   &   8     &  KS 1731--260
\\
    7  & PSR B2334+61 &  9   &  1M 1716--315
\\
    8  & PSR B0656+14  &  10  &  4U 1730--22
\\
    9  &   PSR B0633+1748  &  11  &  4U 2129+47
\\
       & (Geminga)  &  12   &  Terzan 5
\\
    10   & PSR B1055--52  &  13   &  SAX J1808.4--3658
\\
    11   & RX J1856.4--3754 & 14   &  XTE J1751--305
\\
    12   & PSR J2043+2740 &  15   &  XTE J1814--338
\\
    13   & RX J0720.4--3125 & 16     &  EXO 1747--214
 \\
    14   & PSR J1741--2054  &  17     &  Terzan 1
\\
    15   &   XMMU J1732--3445  &  18   &  XTE 2123--058
\\
    16   &   Cas A neutron star     &  19   &   SAX J1810.8--2609
\\
   17    &  PSR J0357+3205  &  20   &  1H 1905+000
\\
         &(Morla)&   21   &  2S 1803--45
\\
    18   &  PSR B0531+21 (Crab)   & 22   &  XTE J0929--314
\\
   19   &  PSR J0205+6449      &   23   &  XTE J1807--294
\\
         &(in 3C 58)   & 24   &  NGC 6440 X-2
\\
    %\bottomrule
    \hline
  \end{tabular}
\label{tab:observa}
\end{table}
\setlength{\tabcolsep}{6pt}
\renewcommand{\arraystretch}{1.0}
%%%%%%%%%%%%%%%%%%%%%%%%%%%

\section{Qualitative analysis}
\label{sec:DurcaMass}

We adopt two EOSs, (i) the HHJ EOS used in Paper I and (ii) the BSk21 EOS (\citealt{GCP10, PCGD12,Potekhin_etal13}). The maximum-mass and direct-Urca-onset neutron star models for these EOSs are presented in Table \ref{tab:models}. The important parameter here is the formal (determined by a given EOS) mass density threshold $\rho_{\rm D0}$ for the onset of the electron direct Urca process and the associated minimum mass of the star $M_{\rm D0}$ in which the direct Urca occurs. As seen from Table \ref{tab:models}, $\rho_{\mathrm{D0}}^{\mathrm{HHJ}}=1.00\times10^{15}$ g cm$^{-3}$, $M_{\mathrm{D0}}^{\mathrm{HHJ}}=1.72$ M$\odot$ and $\rho_{\mathrm{D0}}^{\mathrm{BSk21}}=8.21\times10^{14}$ g cm$^{-3}$, $M_{\mathrm{D0}}^{\mathrm{BSk21}}=1.59$ M$\odot$. Note that similar table 3 of Paper I contains two minor typos which do not affect the results of Paper I because calculations were made with correct parameters.

As in Paper I, the neutrino emissivity $Q$ of the electron direct Urca process is presented as $Q=Q_0b$, where $Q_0$ is the basic direct Urca emissivity disregarding sharp (step-like) threshold at $\rho=\rho_{\rm D0}$ \citep{LPPH91}; $b$ is a phenomenological function to shift and broaden the threshold. It is chosen as 
\begin{equation}
    b(x,\xi) = \begin{cases}
                        0.5\left[1+\erf{\left(x-\xi\omega\right)}\right] &
                        {\rm at}~-\omega \leq x-\xi\omega \leq \omega, \\
                        0 &{\rm at}~ x-\xi\omega < -\omega, \\
                        1 &{\rm at}~ x-\xi\omega > \omega.
                    \end{cases}
\label{e:broadD}
\end{equation}
Here, erf($x$) is the error function; $x=(\rho-\rho_{\mathrm{D0}})/(\alpha \rho_{\mathrm{D0}})$, $\alpha\sim \Delta\rho_{\rm D}/\rho_{\rm D0}$ being the threshold broadening factor and $\Delta\rho_{\rm D}$ a characteristic broadening density interval; $\xi$ (with $-1\! \leq\! \xi\! \leq\! 1$) specifies the position of the actual electron direct Urca 
threshold, $\rho=\rho_{\mathrm{D}}$ (Section \ref{sec:intro}). In fact, $b$ is determined by two physical quantities, $\rho_{\rm D}$ and $\alpha$. At $\xi=0$ the threshold is not shifted from the `basic' value $\rho_{\rm D}=\rho_{\mathrm{D0}}$ prescribed by EOS. If $\alpha \to 0$, the function $b$ becomes step-like with the jump at $\rho=\rho_{\rm D}$. If $\xi \neq 0$, the threshold $\rho_{\mathrm{D}}$ is shifted,
\begin{equation}
    \rho_{\mathrm{D}}=\rho_{\mathrm{D0}}\left(1+\alpha \xi \omega\right),
\label{e:rhoD}
\end{equation}
where $\omega \approx 5.40988$ is an auxiliary quantity which defines the cut-off level $10^{-14}$ of the broadening function, $0.5\left[1-\erf(\omega)\right]=10^{-14}$.
At $\rho=\rho_{\rm D}$ the broadening function $b=0.5$; the function 
varies from 0.1 to 0.9 in the density interval $-0.91\,\alpha \leq (\rho-\rho_{\rm D})/\rho_{\rm D} \leq 0.91 \,\alpha$. The shape of the broadening function is actually the same as in Paper I but the threshold position is allowed to be shifted from $\rho_{\rm D0}$. The muon direct Urca threshold is shifted and broadened in the same way as the electron one which seems rather unimportant. Note that the shape of our broadening factor $b$ as a function of $\rho$ is symmetric with respect to $\rho=\rho_{\rm D}$ (fig.\ 8 of Paper I). In reality, it can be asymmetric and the asymmetry may affect evolution of neutron stars which remains to be studied in the future.
%%%%%%%%%%%%%%%%%%%%%%%%%%%%%%%%%%
\renewcommand{\arraystretch}{1.10}
\setlength{\tabcolsep}{4.6pt}
\begin{table}
\caption{Gravitational masses $M$, central densities $\rho_{c14}$ (in units of $10^{14}$ g~cm$^{-3}$) and circumferential radii $R$ of neutron star models with the HHJ and BSk21 EOSs.}
\centering
\begin{tabular}{ c l l l l l l }
\toprule
\multirow{2}{*}{EOS / model}  &  \multicolumn{3}{c}{HHJ} & \multicolumn{3}{c}{BSk21} \\
\cmidrule(lr){2-4}
\cmidrule(lr){5-7}
&  $M/{\rm M\odot}$  &  $\rho_{c14}$  &  $R$ (km)  &  $M/{\rm M\odot}$  &  $\rho_{c14}$   &  $R$ (km)  \\
\midrule
Maximum mass    & 2.16  &  24.5  & 10.84 & 2.27  &  22.9  &  11.04     \\
Direct Urca onset & 1.72  &  10.0  & 12.49  & 1.59  &  8.21  &  12.59    \\
 \hline
\end{tabular}
\label{tab:models}
\end{table}
\setlength{\tabcolsep}{6pt}
\renewcommand{\arraystretch}{1.0}
%%%%%%%%%%%%%%%%%%%%%%%%%%%%%%%%%%

Let us perform a qualitative analysis of the effects of the direct Urca process in neutron stars. These effects  are characterized by two parameters, $\rho_{\rm D}$ and $\alpha$. In addition, we introduce minimum and maximum masses of neutron stars involved in our analysis, $M^*_{\rm min}$ and $M^*_{\rm max}$. They may be not the absolute minimum and maximum masses of neutron stars (e.g., \citealt{HPY07}) but rather minimum and maximum masses in the ensemble of INSs {\it and} accreting neutron stars in XRTs we would like to study. The masses of neutron stars in XRTs should be overall higher because of accretion. Therefore, $M^*_{\rm max}$ refers to XRTs.

By varying $\rho_{\rm D}$ and $\alpha$, we can estimate upper and lower values of $\rho_{\mathrm{D}}$ ($\rho_{\mathrm{D\,max}}$ and $\rho_{\mathrm{D\,min}}$) and associated upper and lower masses ($M_{\mathrm{D\,max}}$ and $M_{\mathrm{D\,min}}$) at which the direct Urca process appears. If $\rho_{\mathrm{D}}$ is too low, the direct Urca is allowed in low-mass stars making them cooler. If, on the other hand, $\rho_{\mathrm{D}}$ is too high, the direct Urca is suppressed in high-mass stars making them warmer.

Both values, $M_{\mathrm{D\,max}}$ and $M_{\mathrm{D\,min}}$, can be estimated by comparing theoretical cooling/heating curves with observational data. They depend on the broadening factor $\alpha$ and on assumed values of $M^*_{\rm min}$ and $M^*_{\rm max}$. Our formal procedure to estimate possible ranges of $\alpha$ and $\rho_{\rm D}$ (or $M_{\rm D}$) is like this. First, too small broadening ($\alpha \lesssim 0.05$) does not allow us to explain observations of many isolated and accreting neutron stars  (fig 9--12 of Paper I). With this in mind we consider the case of $\alpha \gtrsim 0.05$. For each $\alpha$ we calculate warmest cooling and heating curves (taking $M=M^*_{\rm min}$) at different values of $\rho_{\rm D}$. The minimum value $\rho_{\mathrm{D\,min}}$ is determined as the one which shifts the
warmest cooling or heating curve (the values of $T_\mathrm{s}^\infty$ or $L_\gamma^\infty$) down by 5\% with respect to the curve calculated for non-broadened direct Urca threshold. Of course, the assumed 5\% level is conditional and can be changed but with qualitatively the same results. It is important that this procedure gives nearly the same $\rho_{\rm D\,min}$ (or $M_{\rm D\,min}$) for INSs and XRTs.

As mentioned above, the coldest neutron stars ($M=M^*_{\rm max}$) are thought to be in XRTs. Their heating curves can also be calculated for different values of $\rho_{\rm D}$ at each $\alpha\geq 0.05$. We estimate $\rho_{\mathrm{D\,max}}$ (or $M_{\rm D\,max}$) as such which shifts the coldest heating curve up by 5\%. This estimate is almost insensitive to $\alpha$.
%%%%%%%%%%%%%%%%%%%%%%%%%%%
\begin{figure}
\centering
\includegraphics[width=0.475\textwidth]{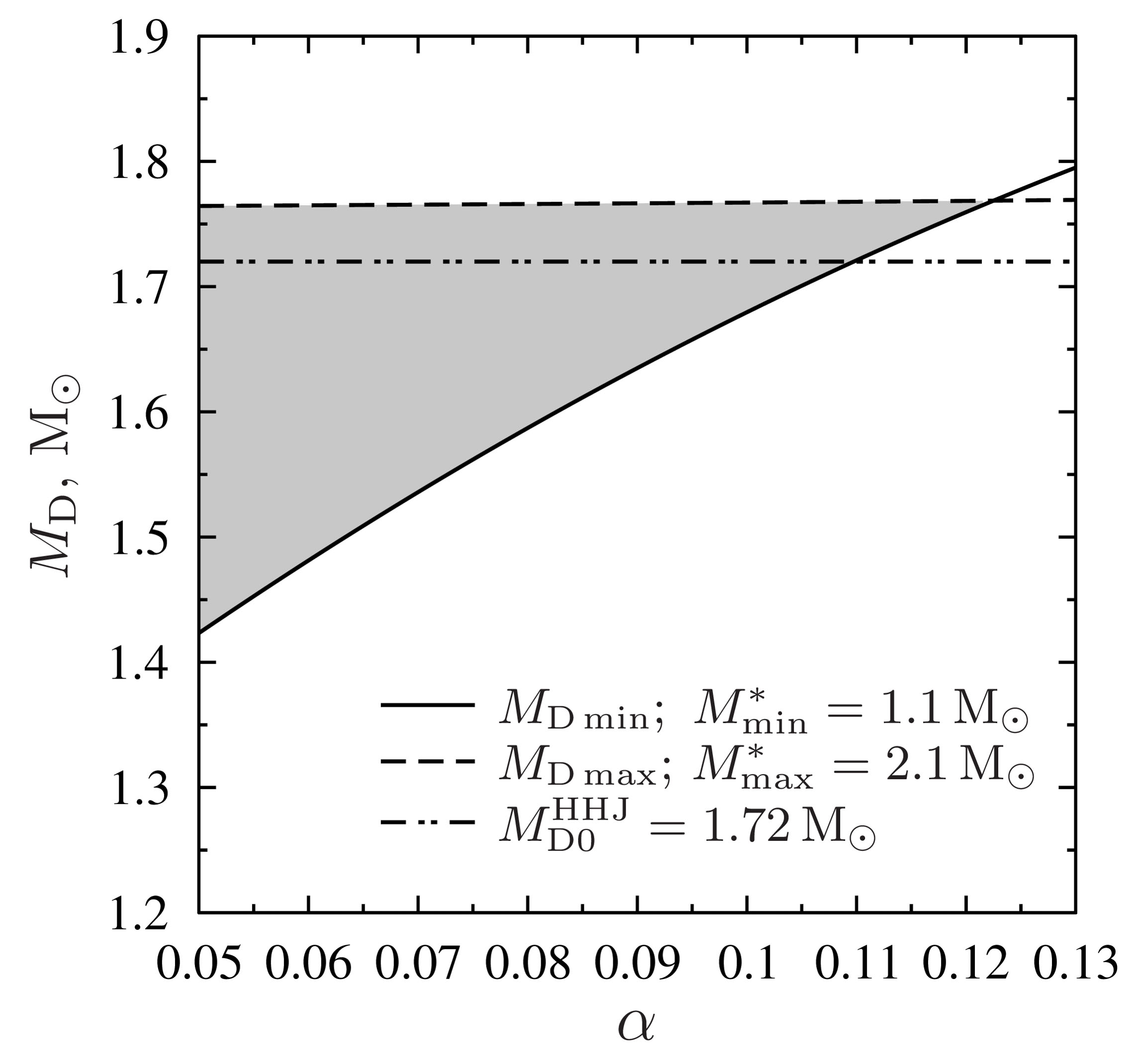}
\caption{Ranges of possible values of $M_{\rm D}$ versus $\alpha$ (shaded area) for the HHJ EOS at $M^*_{\rm min}=1.1\,{\rm M}\odot$ and $M^*_{\rm max}=2.1\,{\rm M}\odot$. The dashed line shows $M_{\rm D\,max}$ while the solid line is $M_{\rm D\,min}$. The double dot--dashed line is $M_{\rm D0}^\mathrm{HHJ}$ (Table \ref{tab:models}). See text for details.}
\label{fig:HHJ-D-Mass}
\end{figure}
\begin{figure}
\centering
\includegraphics[width=0.475\textwidth]{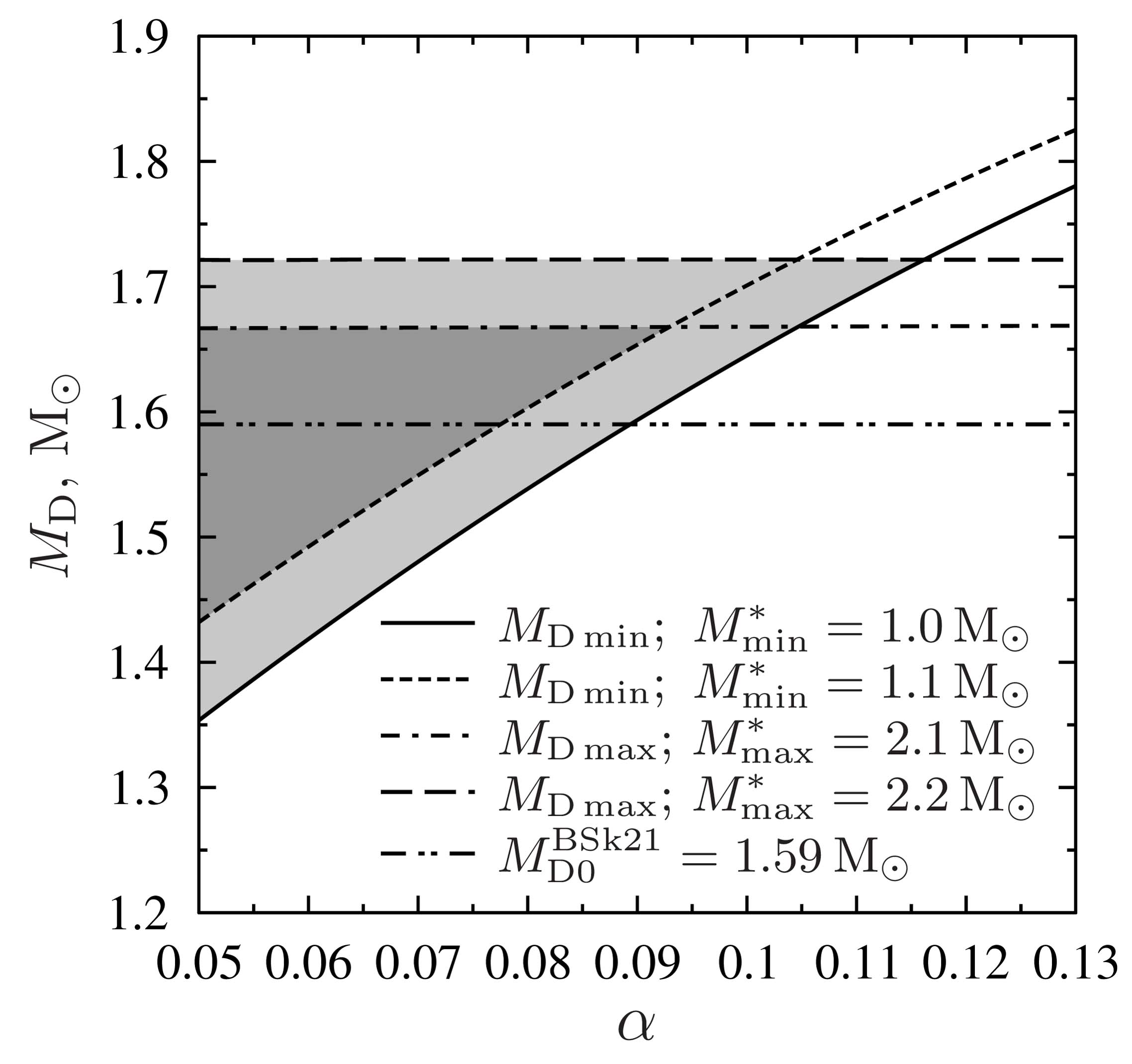}
\caption{Allowable minimum (short-dashed and solid lines) and maximum (long-dashed and dash--dotted lines) masses $M_{\rm D}$ of neutron stars versus $\alpha$ for the BSk21 EOS at $M^*_{\mathrm{min}}=1.0$ and $1.1\,{\rm M}\odot$ and $M^*_{\mathrm{max}}=2.1$ and $2.2\,{\rm M}\odot$. The entire shaded area shows the range of possible values $M_{\rm D}$ for $M_{\rm min}^*=1.0\,{\rm M}\odot$ and $M_{\rm max}^*=2.2\,{\rm M}\odot$, while the denser shading is the same for $M_{\rm min}^*=1.1\,{\rm M}\odot$ and $M_{\rm max}^*=2.1\,{\rm M}\odot$. See text for details.}
\label{fig:BSk21-D-Mass}
\end{figure}
\begin{figure}
\centering
\includegraphics[width=0.475\textwidth]{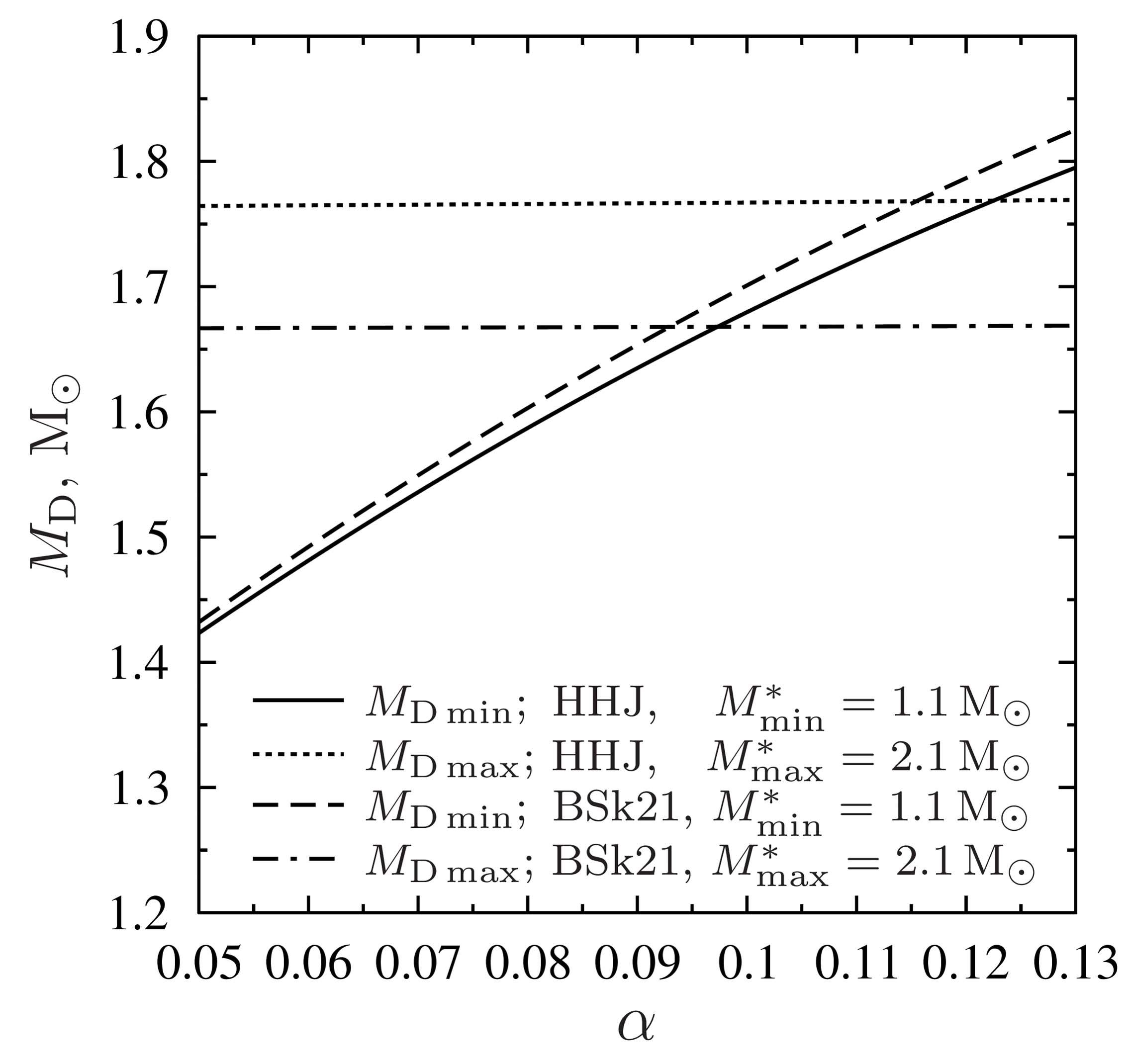}
\caption{Comparison of minimum and maximum allowable masses $M_{\rm D}$ versus $\alpha$ for the HHJ and BSk21 EOSs at $M^*_{\mathrm{min}}=1.1\,{\rm M}\odot$ and $M^*_{\mathrm{max}}=2.1\,{\rm M}\odot$. See text for details.}
\label{fig:HHJ-BSk21-D-Mass}
\end{figure}
%%%%%%%%%%%%%%%%%%%%%%%%%%%%%

The results of this analysis are presented in Figs.\ \ref{fig:HHJ-D-Mass} -- \ref{fig:HHJ-BSk21-D-Mass}. Fig.\ \ref{fig:HHJ-D-Mass} shows allowable values of $M_{\rm D}$ and $\alpha$ (shaded domain) for the HHJ EOS assuming $M^*_{\rm min}=1.1\,{\rm M}\odot$ and $M^*_{\rm max}=2.1\,{\rm M}\odot$ at $\alpha \gtrsim 0.05$. The solid line is $M_{\rm D\,min}$ and the
dashed line is $M_{\rm D\,max}$. The double dot--dashed line is $M_{\rm D0}$ (Table \ref{tab:observa}).

Fig.\ \ref{fig:BSk21-D-Mass} presents $M_{\rm D\,min}$ and $M_{\rm D\,max}$ as a function of $\alpha$ for the BSk21 EOS. The solid and short-dashed lines are $M_{\rm D\,min}$ for $M^*_{\rm min}=1.0$ and $1.1\,{\rm M}\odot$, respectively. The dot--dashed and long-dashed lines show $M_{\rm D\,max}$ for $M^*_{\rm max}=2.1$ and $2.2\,{\rm M}\odot$, respectively. Light-greyed and dark-greyed areas (combined together) show the wide range of allowed values of $M_\mathrm{D}$ corresponding to the mass range of $1.0--2.2$\,M$\odot$, while dark-greyed area (alone) shows narrower range of allowed $M_\mathrm{D}$ values corresponding to the $1.1--2.1$\,M$\odot$
mass range.

Fig.\  \ref{fig:HHJ-BSk21-D-Mass} compares minimum (solid and long-dashed curves) and maximum (dotted and dot-dashed curves) allowable masses $M_{\rm D}$ versus $\alpha$ for the HHJ and BSk21 EOSs at $M^*_{\mathrm{min}}=1.1\,{\rm M}\odot$ and $M^*_{\mathrm{max}}=2.1\,{\rm M}\odot$.

If $M_{\mathrm{D\,min}}$ formally exceeds $M_{\mathrm{D\,max}}$, the broadening of the direct Urca threshold becomes too strong and theoretical cooling/heating curves cannot explain all
observational data. The presented analysis gives a tighter upper constraint on $\alpha$ than in Paper I. As seen from Figs.\ \ref{fig:HHJ-D-Mass} -- \ref{fig:HHJ-BSk21-D-Mass}, $\alpha$ should not exceed $0.12--0.15$.

Also notice that the derived constraints on $M_{\rm D}$ depend on assumed values of $M^*_{\mathrm{min}}$ and $M^*_{\mathrm{max}}$ (which restrict masses of neutron stars of our interest). We have taken $M^*_{\mathrm{min}} = 1.0, 1.1\,\mathrm{M}\odot$ and $M^*_{\mathrm{max}} = 2.1, 2.2\,\mathrm{M}\odot$ which do not contradict observational data. In the next section we use $M^*_{\mathrm{min}} = 1.1\,\mathrm{M}\odot$ and $M^*_{\mathrm{max}} = 2.1\,\mathrm{M}\odot$ to be consistent with Paper I.

%%%%%%%%%%%%%%%%%%%%%%%%%%%%%%%%%%%%%%%%%%%%%%%%%
\section{Position and broadening of direct Urca threshold}
\label{sec:DurcaBroad}

\subsection{Illustrative mass functions}
\label{sec:mfun}

Let us supplement the qualitative analysis of the previous section by some numerical results. We will mainly present probabilities to find isolated and accreting neutron stars in certain regions of
$T_\mathrm{s}^\infty-t$ (for INSs) or $L_\gamma^\infty - \langle \dot{M} \rangle$ (for XRTs) planes compared with observations (Section \ref{sec:intro}, Table \ref{tab:observa}; see also Paper I, tables 1 and 2). Specifically, they are the differential probabilities
\begin{equation}
   \mathrm{d}P_{\rm i}=p_{\rm i}\,(T_{\rm s}^\infty,t)\, \mathrm{d}T_{\rm s}^\infty,
	 \quad
	 \mathrm{d}P_{\rm a}=p_{\rm a}\,(L_\gamma^\infty,\langle \dot{M} \rangle)\,
	 \mathrm{d}L_\gamma^\infty,
\label{eq:probab}
\end{equation}
to find an isolated (i) or accreting (a) neutron star of given age $t$ or mass accretion rate $\langle \dot{M} \rangle$ in small surface temperature or luminosity intervals, $ \mathrm{d}T_{\rm s}^\infty$ or $ \mathrm{d}L_{\gamma}^\infty$, respectively. By construction, the total probabilities $\int p_{\rm i}\, \mathrm{d}T_{\rm s}^\infty$ and $\int p_{\rm a}\,  \mathrm{d}L_\gamma^\infty$ are conserved (independent of $t$ and $\langle \dot{M} \rangle$).

The differential probability densities, $p_{\rm i}\,(T_{\rm s}^\infty,t)$ and $p_{\rm a}\,(L_\gamma^\infty,\langle \dot{M} \rangle)$, will be plotted by greyscaling (in relative units). The denser the scaling, the larger the probability. White regions refer to zero or very low probability. 
A graphical example of the probability distribution is discussed in the end of Section \ref{sec:bestalpha}.

The probabilities are calculated (Paper I) by averaging the families of cooling or heating curves over mass distributions of isolated or accreting neutron stars and over masses of light elements $\Delta M_{\rm le}$ in the heat blanketing envelopes of the stars. For clarity, by dashed lines we will present also the initial cooling or heating `reference' curves for neutron stars with masses $M=$ 1.1, 1.2,\,\ldots, 2.1\,${\rm M}\odot$ with iron heat blankets. Higher curves correspond to lower $M$. Curves for low-mass stars ($M < M_{\rm D}$) often merge in nearly the same curve.

\begin{figure*}
\centering
\includegraphics[width=0.45\textwidth]{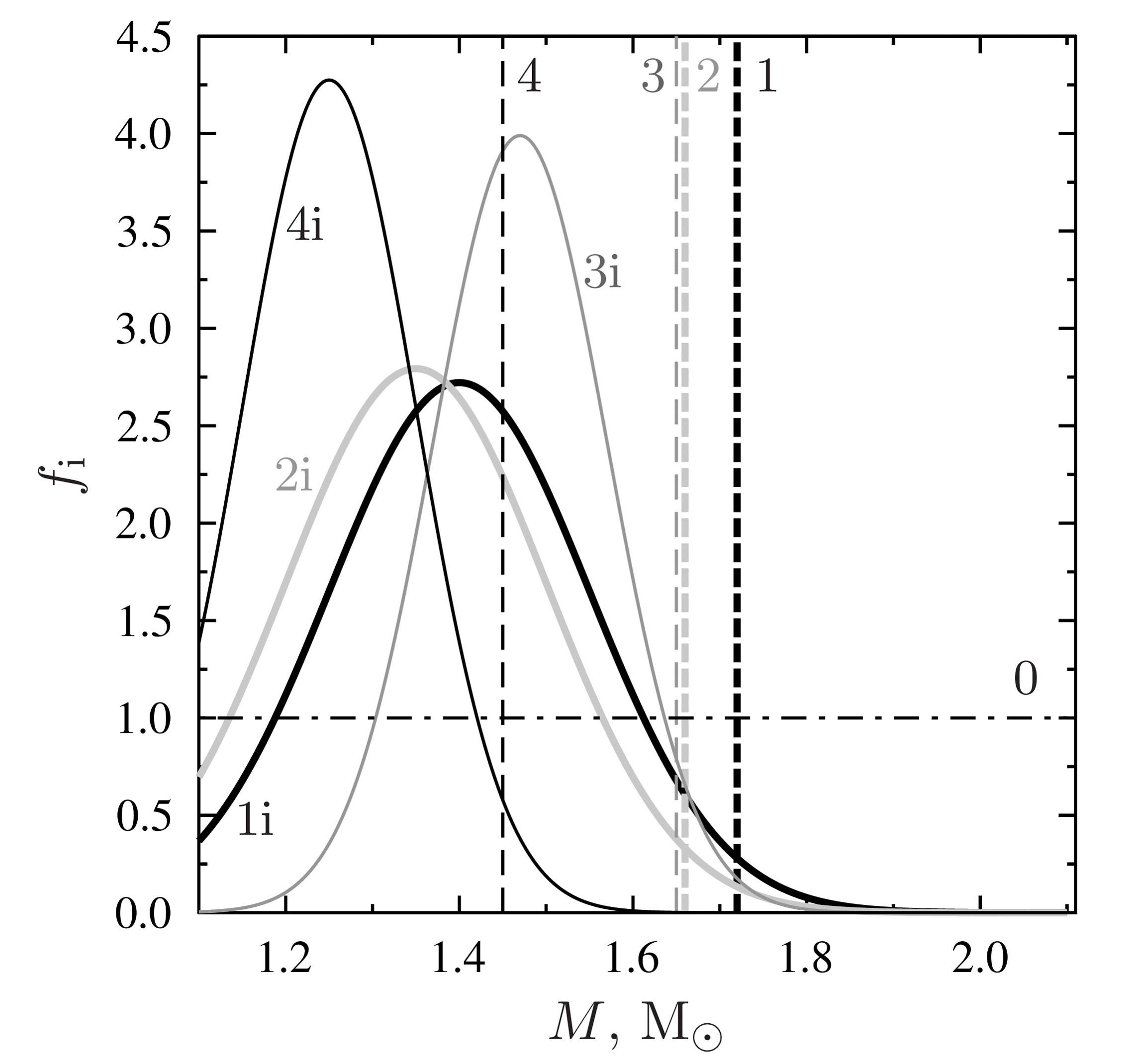}%
\includegraphics[width=0.45\textwidth]{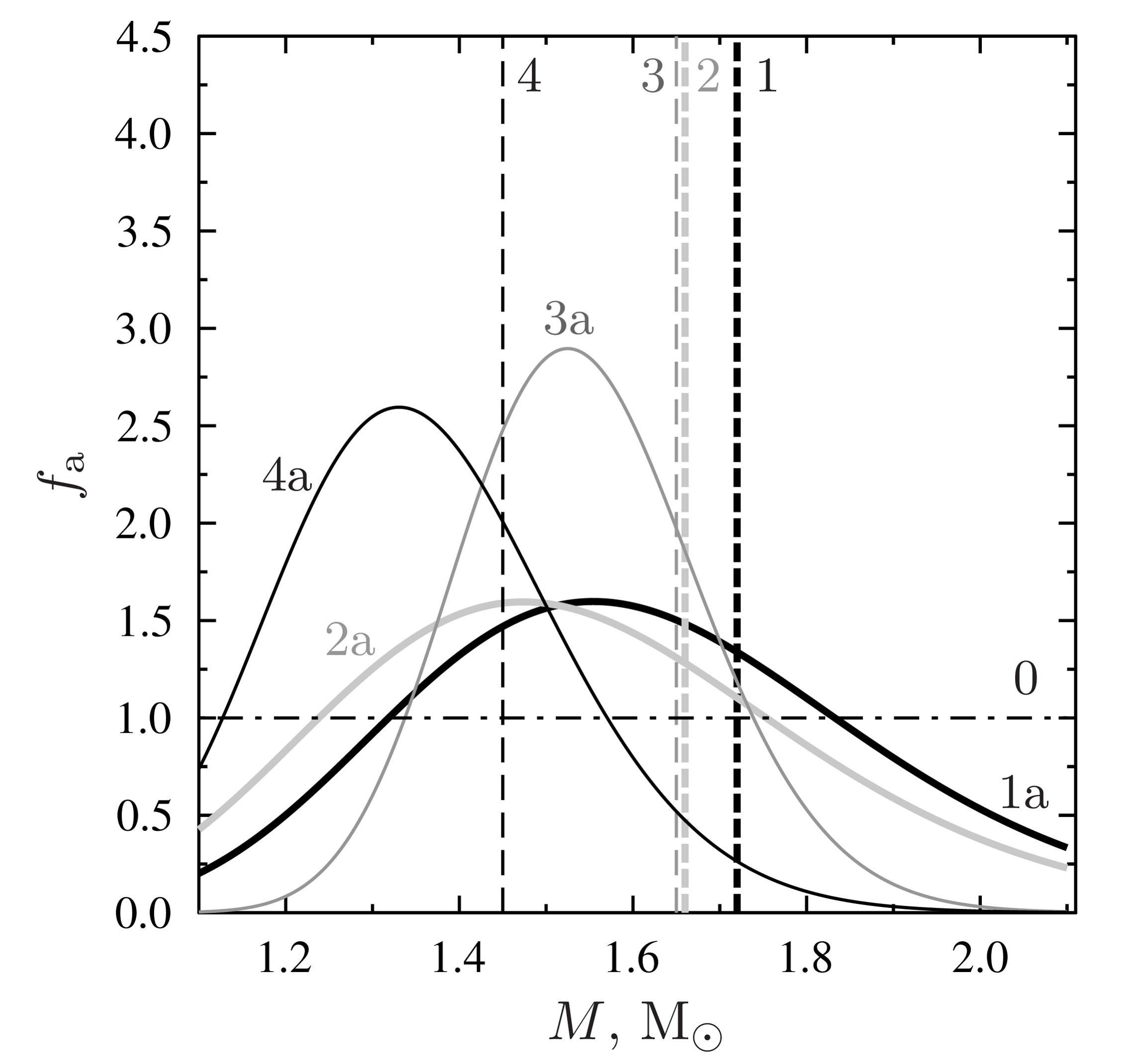}
\caption{Illustrative models 1--4 (Table \ref{tab:mfun}) of mass distribution functions ({\ref{e:mass-distrib}}) for INSs (i, left-hand panel) and accreting neutron stars  in XRTs (a, right-hand panel). The dot--dashed line `0' refers to the uniform mass distribution. Vertical dashed lines show values of $M_\mathrm{D}$ for corresponding models.} 
\label{fig:mfun}
\end{figure*}

As in Paper I, the distribution function over  $\Delta M_{\rm le}$ will be taken uniform, and $\Delta M_{\rm le}/M$ is allowed to vary from 0 to its maximum value $10^{-7}$. Following Paper I we use normal mass distributions $f(M)$ for INSs and lognormal for neutron stars in XRTs,
\begin{equation}
\begin{split}
     &f_\mathrm{i}(M)=\frac{1}{N_\mathrm{i}}\frac{1}{\sqrt{2\uppi}\,
        \sigma_\mathrm{i}}\exp{\left(-\frac{\left(M-\mu_\mathrm{i}\right)^2}
        {2\sigma_\mathrm{i}^2}\right)}, \\
    &f_\mathrm{a}(M)=\frac{1}{N_\mathrm{a}}\frac{1}
    {\sqrt{2\uppi}\,M\sigma_\mathrm{a}}\exp{\left(-\frac{\left(\ln{\left[M/{\rm M}\odot\right]}
    -\mu_\mathrm{a}\right)^2}{2\sigma_\mathrm{a}^2}\right)}.
\end{split}
\label{e:mass-distrib}
\end{equation}
Here `i' stands for INSs and `a' is for XRTs; $\sigma_{\rm i}$, $\mu_{\rm i}$, and $\sigma_{\rm a}$, $\mu_{\rm a}$ are the parameters of these distributions; $N_{\rm i}$ and $N_{\rm a}$ are
normalization factors. We will take four pairs of mass functions 1--4 listed in Table \ref{tab:mfun}. Note that $\sigma_{\rm i}$ and $\mu_{\rm i}$ are expressed in units of ${\rm M}\odot$, while
$\sigma_{\rm a}$ and $\mu_{\rm a}$ are dimensionless. We consider these distributions within the mass range from $M^*_{\rm min}$=1.1\,${\rm M}\odot$ to $M^*_{\rm max}$=2.1\,${\rm M}\odot$, with $f(M)=0$ outside this range. These distributions are plotted in Fig.\ \ref{fig:mfun}. Their parameters have been chosen `by eye' to get better agreement between the theory and observations in each specific case described below. The distributions 1i and 1a were used in Paper I; other distributions are new. Evidently, these distributions are not unique; a detailed study of most suitable distributions will be done in future publications. In all four cases (1--4) the peaks of $f_{\rm a}(M)$ are shifted to higher $M$ (by $\sim 0.15 \,{\rm M}\odot$) with respect to the peaks of $f_{\rm i}(M)$, as a natural result of accretion. By way of illustration, we will also take the uniform mass distribution within the same mass interval from 1.1 to 2.1\,${\rm M}\odot$ for INSs and XRTs (plotted by the dot--dashed curve `0' in Fig.\ \ref{fig:mfun}).

%%%%%%%%%%%%%%%%%%%%%%%%%%%%%%%%%%
\renewcommand{\arraystretch}{1.10}
\begin{table}
\caption{Parameters of model mass distribution functions (\ref{e:mass-distrib}) of INSs (i) and accreting neutron stars (a) in XRTs and corresponding values of $M_\mathrm{D}$ used in illustrative examples. See also Fig.\ \ref{fig:mfun}.}
\centering
\begin{tabular}{ c l l l l l }
\toprule
Model  &  $\mu_{\rm i}$  &  $\sigma_{\rm i}$  &  $\mu_{\rm a}$  &  $\sigma_{\rm a}$  &  $M_\mathrm{D}$  \\
\midrule
1 &  1.40 & 0.15  & 0.47  & 0.17  &   1.72 \\
2 &  1.35 & 0.15  & 0.42  & 0.18  &   1.66 \\
3 &  1.47 & 0.10  & 0.43  & 0.09  &   1.65 \\
4 &  1.25 & 0.10  & 0.30  & 0.12  &   1.45 \\
 \hline
\end{tabular}
\label{tab:mfun}
\end{table}
\setlength{\tabcolsep}{6pt}
\renewcommand{\arraystretch}{1.0}
%%%%%%%%%%%%%%%%%%%%%%%%%%%%%%%%%%

%%%%%%%%%%%%%%%%%%%%%%%%%%%%%%%%%%

\subsection{Very small broadening, $\alpha \lesssim 0.05$}
\label{sec:smallalpha}

In Paper I, we demonstrated the effect of too small broadening, with $\alpha \lesssim 0.05$. It splits the populations of INSs and XRTs into families of  rather warm and cold sources separated by `gaps,' in disagreement with observations (figs 9--12 of Paper I). It was shown for the HHJ EOS. Now we obtain similar results for the BSk21 EOS. Note that at $\alpha \approx 0.05$ the `gap' is narrow and, in principle, could be closed by tuning mass functions. However, to this aim one needs sufficiently sharp and narrow mass distributions which seem unlikely. Therefore, the conclusion that the broadening of the direct Urca threshold is not too small looks solid. At $\alpha \gtrsim 0.05$ the `gaps' disappear and the theory can be consistent with the data.

\subsection{Most `successful' broadening, $\alpha \approx 0.08--0.10$}
\label{sec:bestalpha}

As shown in Paper I (for the HHJ EOS), the case of $\alpha \sim 0.1$ is most suitable to explain the observations. Let us focus on this important case. For a fixed $\alpha \sim 0.1$ one should find most appropriate value of $M_{\rm D}$ and most appropriate mass functions $f_{\rm i}(M)$ and $f_{\rm a}(M)$ to explain the data.

\begin{figure*}
\centering
\includegraphics[width=0.45\textwidth]{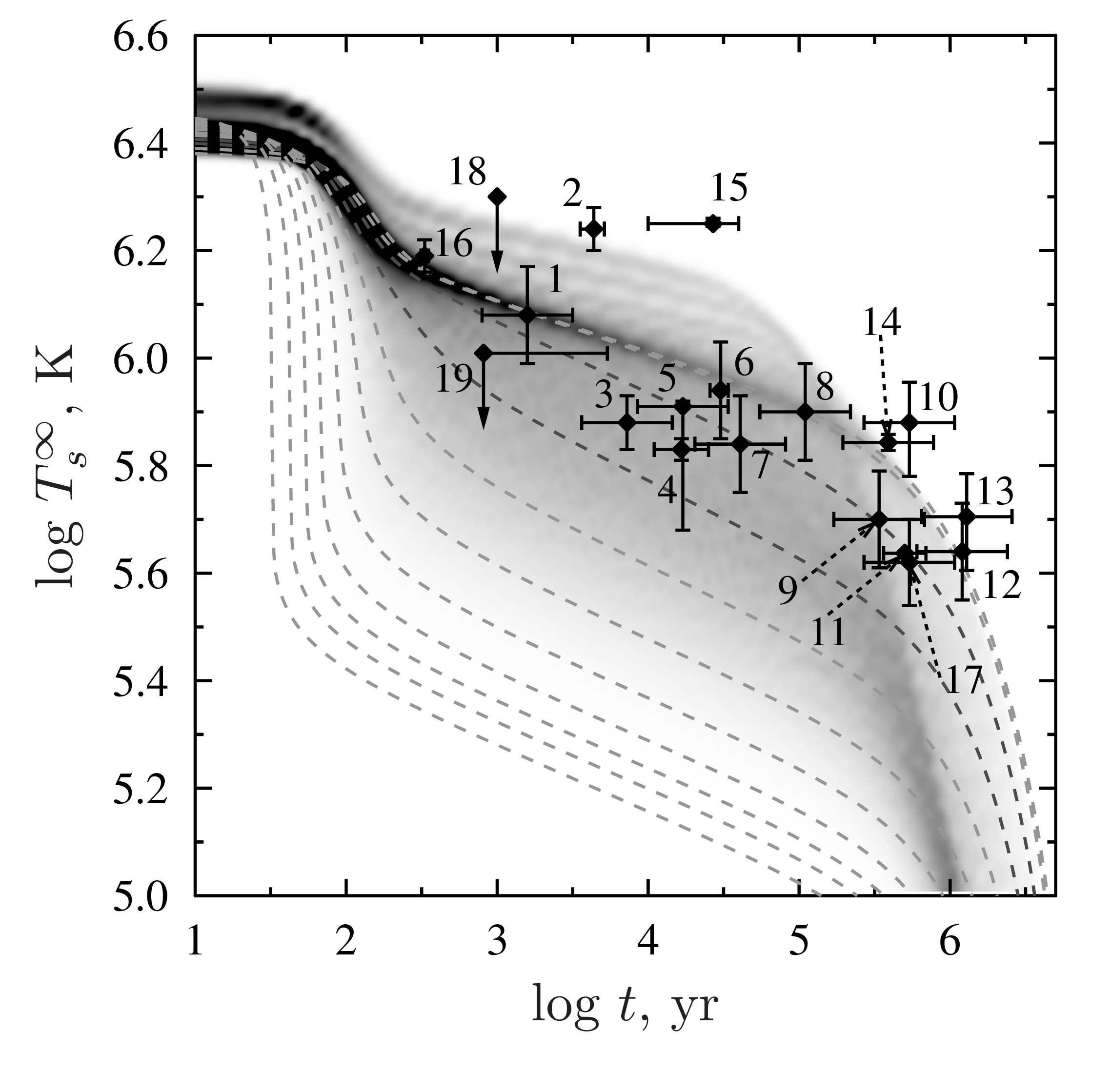}%
\includegraphics[width=0.45\textwidth]{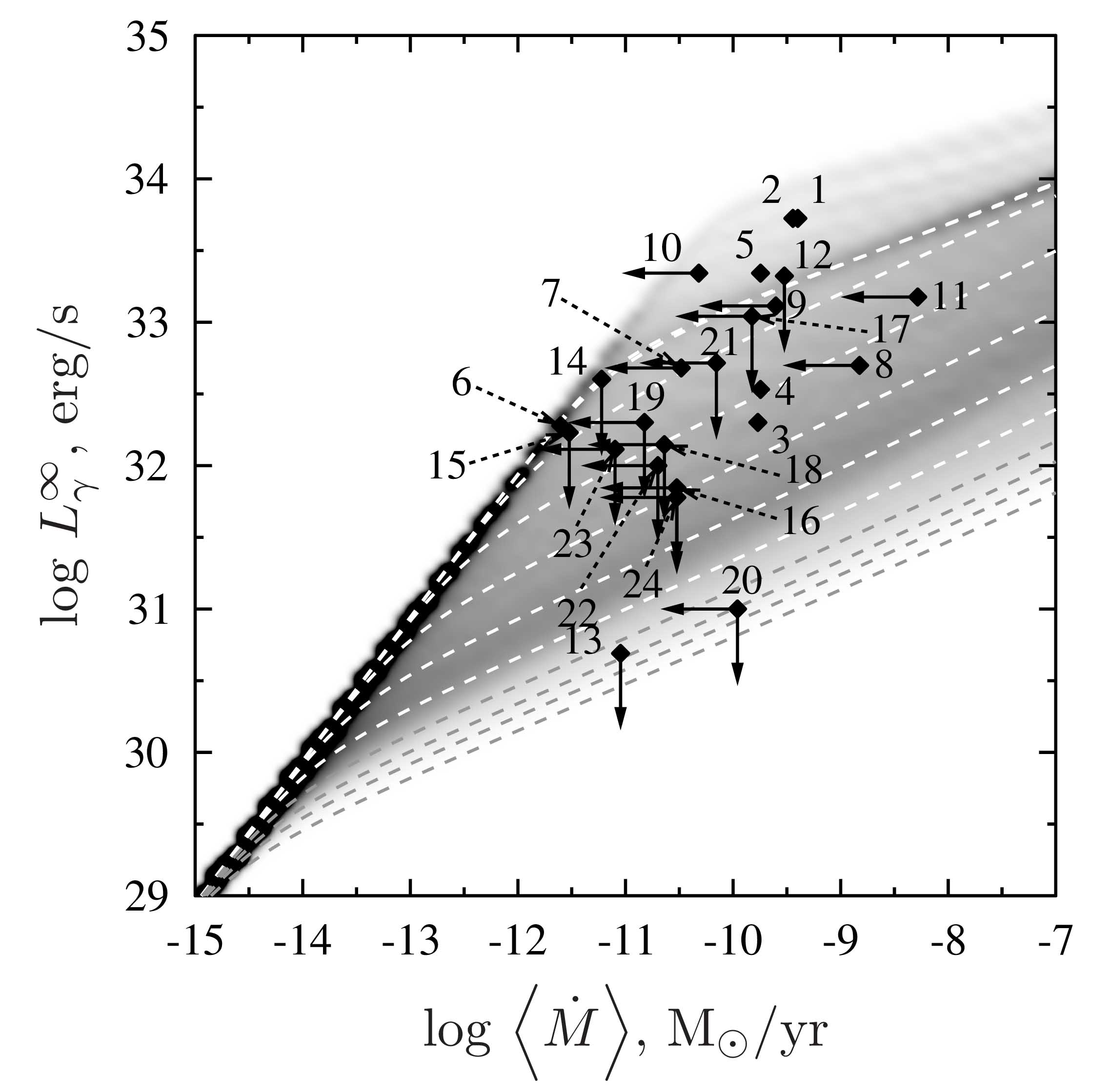}
\caption{Probability distributions for INSs in the $T_\mathrm{s}^\infty-t$ plane (left) and 
for accreting neutron stars in the $L_{\gamma}^\infty-\langle\dot{M} \rangle$ plane (right) compared with observations. The HHJ EOS is used; the distributions of neutron star masses are 1i and 1a, and the distribution over the mass of light elements in surface layers is uniform. Dashed lines show 11 `reference' cooling curves for stars with iron envelopes and masses (from top to bottom) $M=1.1\,, 1.2\,, \ldots,2.1\,\mathrm{M}\odot$. The direct Urca threshold is broadened but not shifted, $\alpha=0.1$, $M_\mathrm{D}=M_{\mathrm{D0}}^\mathrm{HHJ}$. See the text for details.} \label{fig:CoolD_10_HHJ}
\end{figure*}

Fig.\ \ref{fig:CoolD_10_HHJ} shows the probability distributions for INSs (in the $T_\mathrm{s}^\infty-t$ plane, left-hand panel) and XRTs (in the $L_{\gamma}^\infty-\langle\dot{M} \rangle$ plane, right-hand panel) assuming the HHJ EOS, $\alpha=0.1$, $M_\mathrm{D}=M_\mathrm{D0}^\mathrm{HHJ}=1.72\,{\rm M}\odot$, and the mass functions 1i and 1a (after figs 13 and 14 of Paper I). This model explains the observations reasonably well. According to Fig.\ \ref{fig:HHJ-D-Mass}, we can slightly vary $M_\mathrm{D}$
around $M_\mathrm{D0}$ and slightly adjust the mass functions. These new explanations will also be satisfactory although the explanation of Paper I is one of the best. The source 15 on the left-hand panel (XMMU J1732--3445, Table \ref{tab:observa}; \citealt{Klochkov_etal13}) is too hot to be explained by the present model. As shown in \cite{Klochkov_etal15} and Paper I, it can be interpreted as a neutron star with strong proton superfluidity in the core and heat blanketing envelope fully made of light elements.
The neutron star in Pup A (source 2) also seems to require proton superfluidity and the heat blanket of light elements but the requirement is much less stringent. We do not include the effects of strong proton superfluidity here to simplify our analysis. 

\begin{figure*}
\centering
\includegraphics[width=0.45\textwidth]{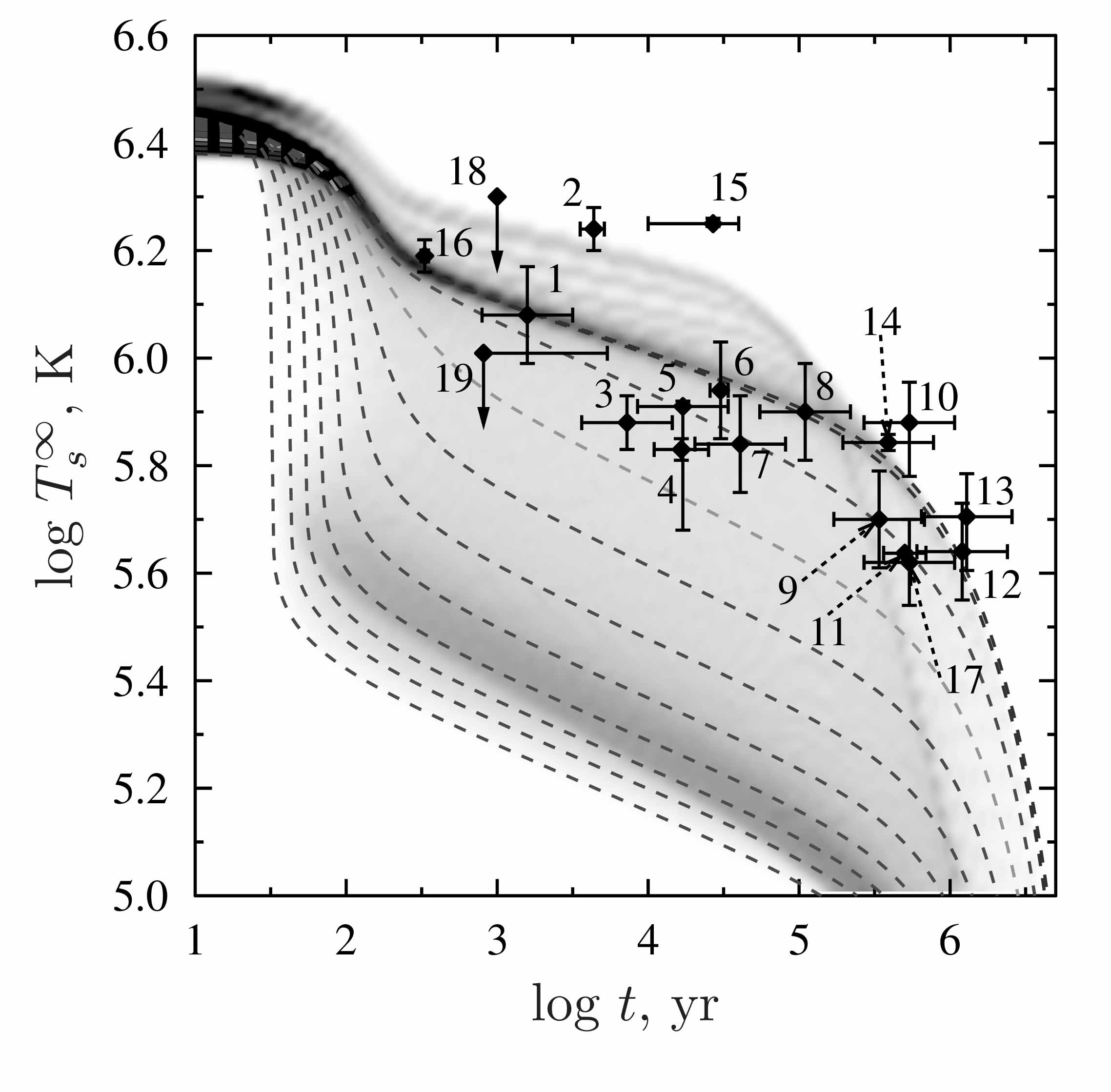}%
\includegraphics[width=0.45\textwidth]{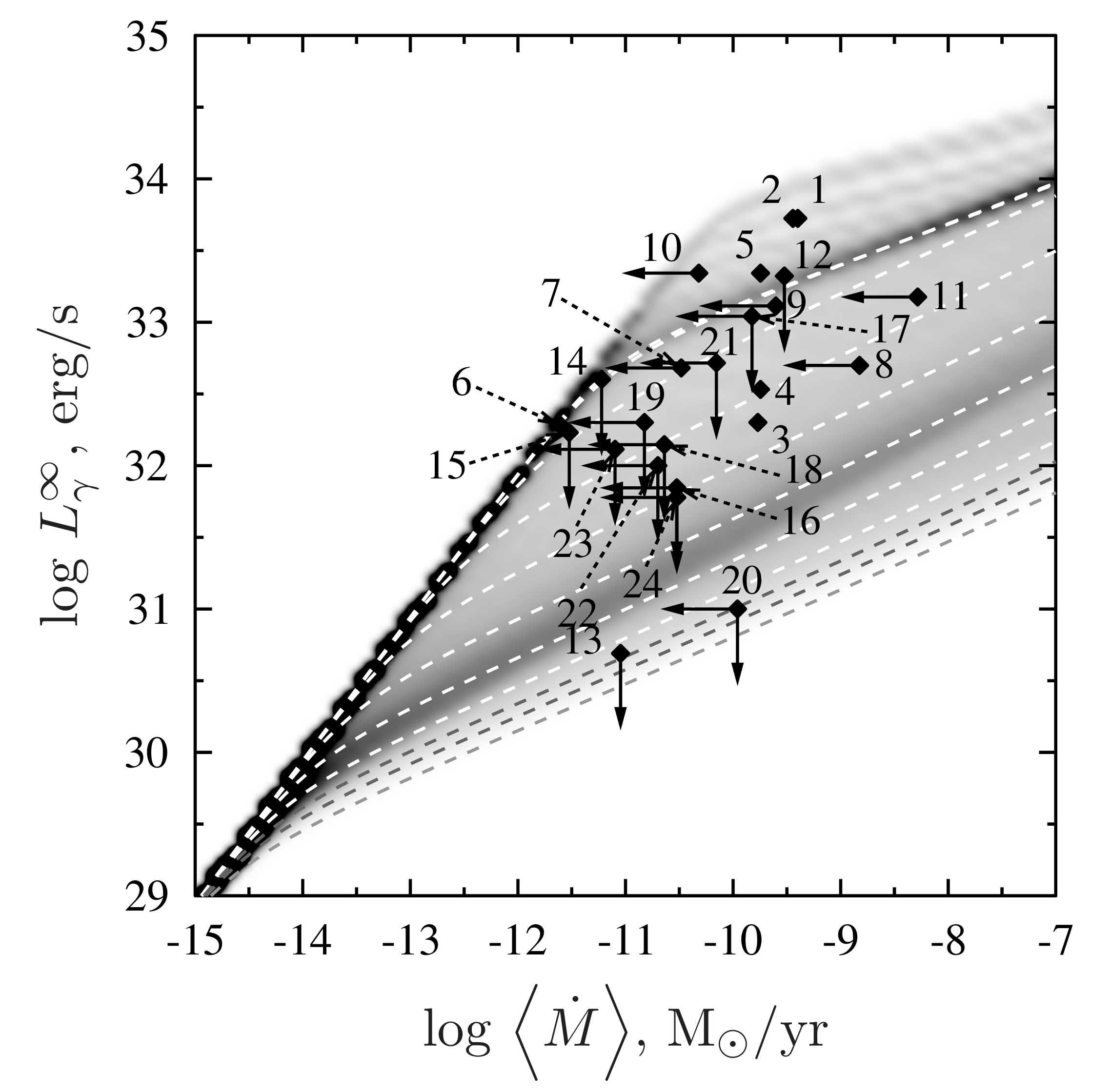}
\caption{Same as in Fig.\ \ref{fig:CoolD_10_HHJ} but for the uniform mass distributions. See the text for details.}
\label{fig:CoolD_10u_HHJ}
\end{figure*}

For illustration, Fig.\ \ref{fig:CoolD_10u_HHJ} presents also the probability distributions for isolated and accreting neutron stars with the HHJ EOS but using the uniform mass distribution `0' instead of 1i and 1a. Such a model was not studied in Paper I.  In contrast to the previous one (Fig.\ \ref{fig:CoolD_10_HHJ}), it predicts the existence of cold INSs which have not been observed. In addition, it is less successful in explaining the existence of many neutron stars in XRTs which are intermediate between coldest and hottest ones. The presented example demonstrates the importance of mass distribution functions $f(M)$ in our analysis.

\begin{figure*}
\centering
\includegraphics[width=0.45\textwidth]{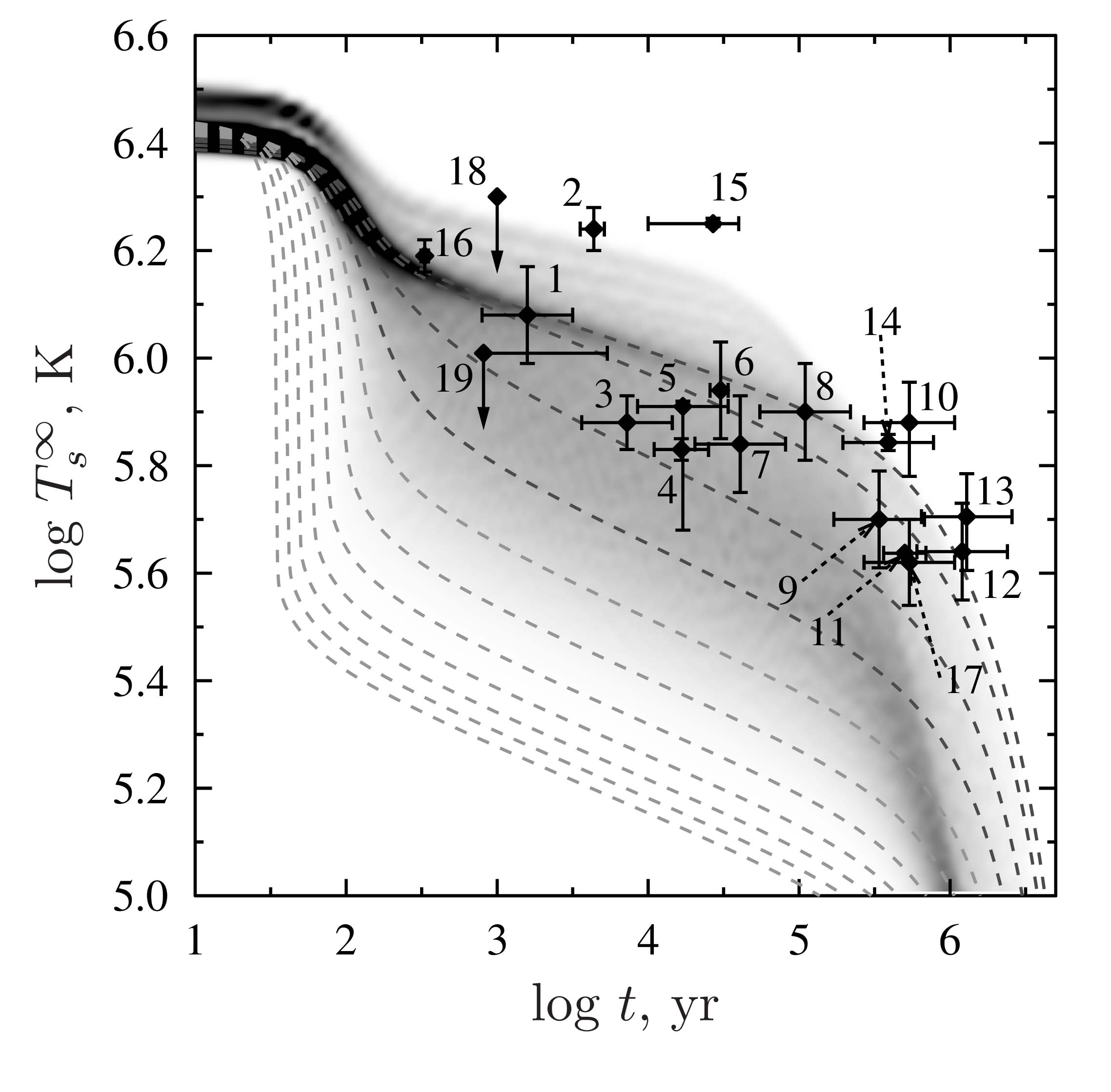}%
\includegraphics[width=0.45\textwidth]{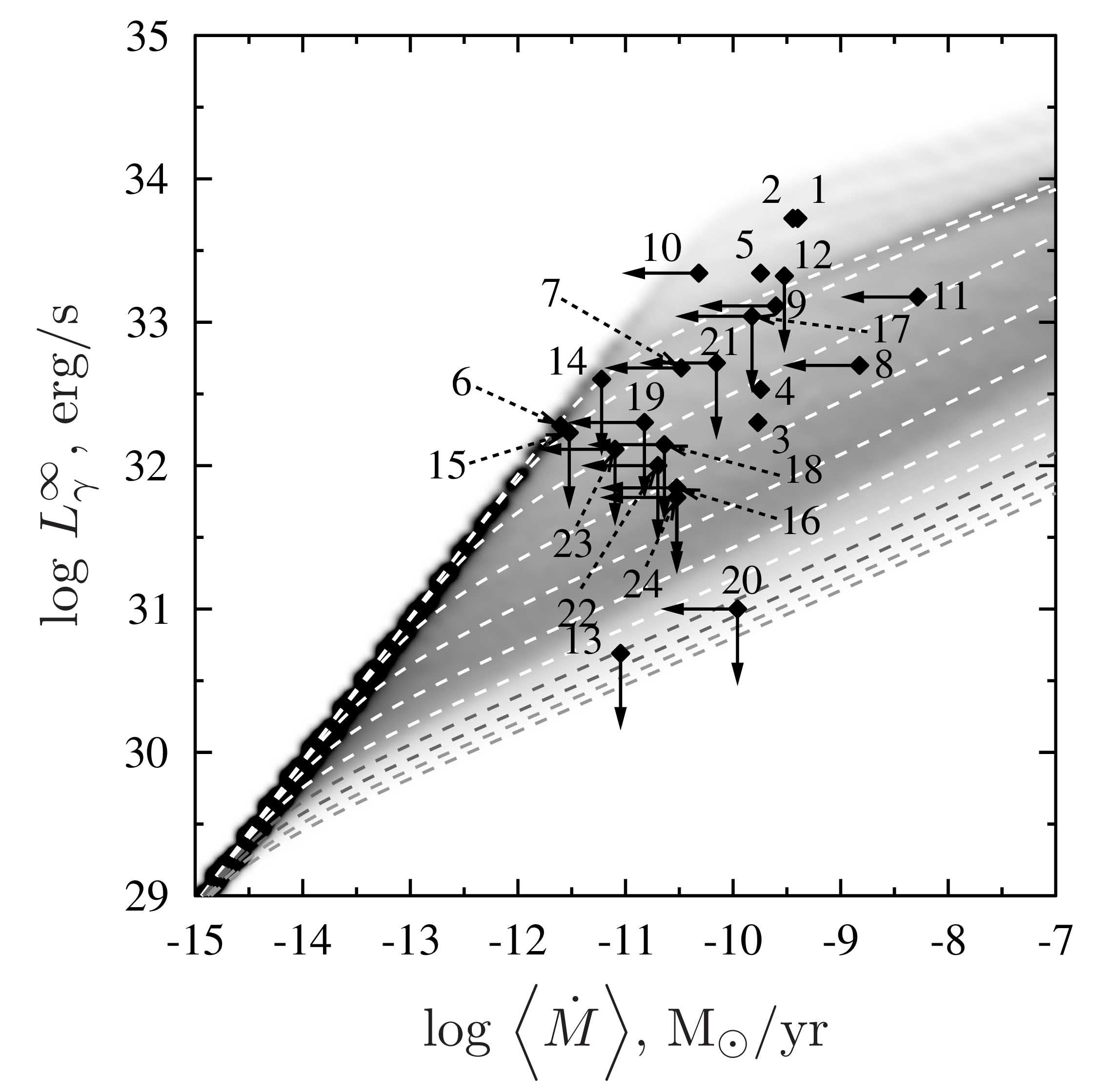}
\caption{Same as in Fig.\ \ref{fig:CoolD_10_HHJ} but for the BSk21 EOS, $\alpha=0.09$, $M_\mathrm{D}=1.66$ M$\odot$, and neutron star mass distributions 2i and 2a.} \label{fig:CoolD_11_BSk21}
\end{figure*}

Now let us consider the BSk21 EOS, whose formal direct Urca threshold corresponds to $M_{\rm D0}=1.59\,{\rm M}\odot$. As seen from Fig.\ \ref{fig:BSk21-D-Mass}, the values $\alpha=0.1$ and $M_{\rm D}=M_{\rm D0}^{\rm BSk}$ are not most suitable for explaining the observations. After several trial calculations we have found that the case of $\alpha=0.09$ and $M_{\rm D}=1.66\,{\rm M}\odot$, where the actual direct Urca threshold is shifted to higher densities, is more successful. The distributions of isolated and accreting neutron stars over respective diagrams for this case are plotted in Fig.\ \ref{fig:CoolD_11_BSk21}. The mass distributions 2i and 2a (Fig.\ \ref{fig:mfun}) for this case are also modified with respect to 1i and 1a to achieve better agreement with the data. The modified mass distributions contain slightly less massive isolated and accreting neutron stars. Otherwise, theoretical models presented in Figs.\ \ref{fig:CoolD_10_HHJ} and \ref{fig:CoolD_11_BSk21} look rather similar. Note that sufficiently large masses $M_{\rm D}\gtrsim 1.5 \,{\rm M}_{\odot}$ agree with some theoretical expectations (e.g., \citealt{Blaschke_etal06}).

\begin{figure}
\centering
\includegraphics[width=0.475\textwidth]{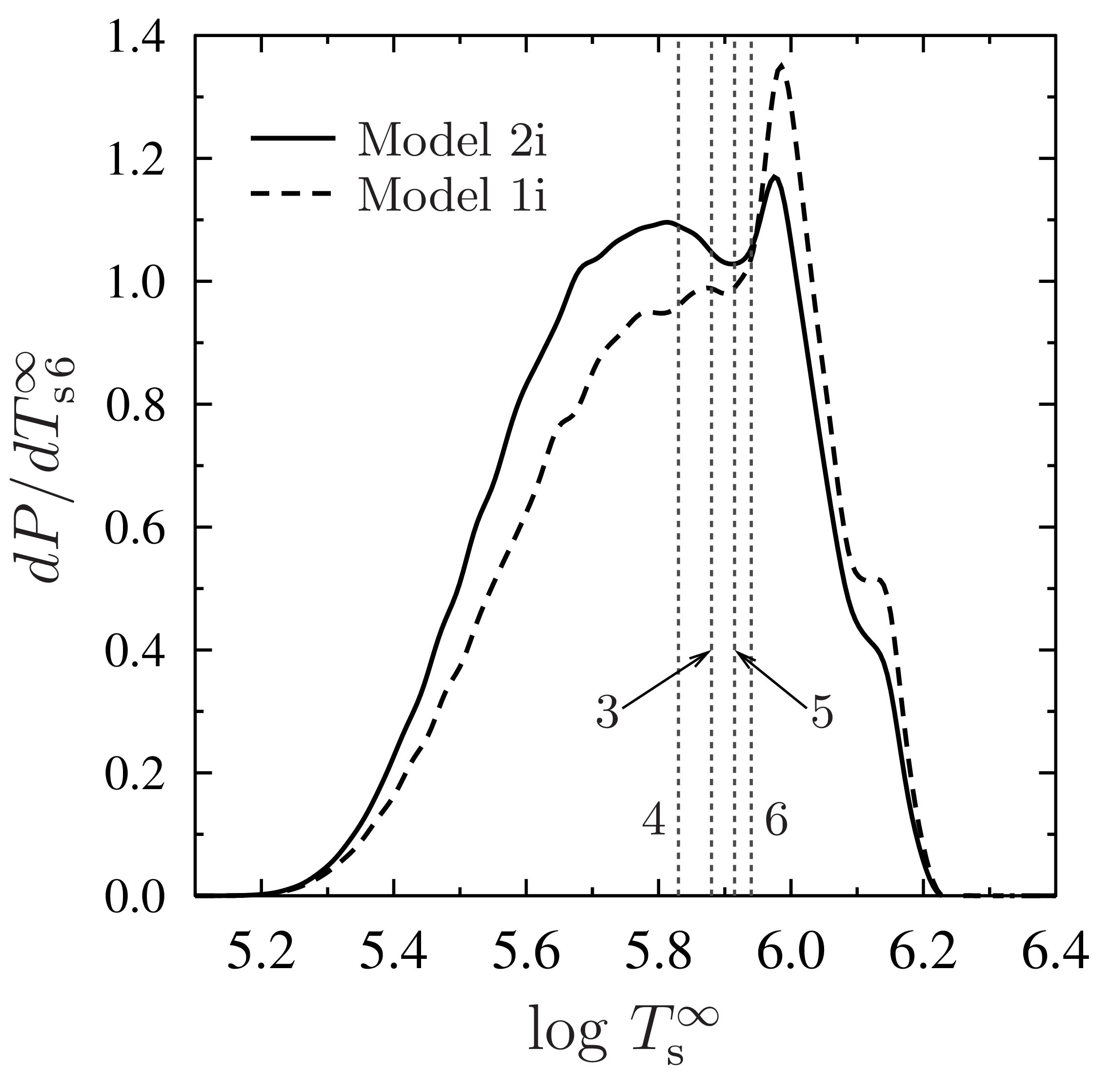}
\caption{Theoretical differential probability distributions $dP/dT_{\rm s\,6}^{\infty}$
($T_{\rm s\,6}^\infty=T_{\rm s}^\infty/10^6~{\rm K}$) to observe an INS as functions of $T_{\rm s}^\infty$. The probability is averaged over the age interval from $t=10^{3.5}$ to $10^{4.5}$ yr. The dashed line is calculated for the HHJ EOS with the mass distribution 1i, while the solid line is for the BSk21 EOS and the mass distribution 2i. Vertical dotted lines show the central surface temperatures for four INSs, sources 3--6. See the text for details.}
\label{fig:Probab}
\end{figure}

To visualize the theoretical probabilities \eqref{eq:probab}, in Fig.\ \ref{fig:Probab} we show the calculated differential probability distributions $\mathrm{d}P/\mathrm{d}T_{\rm s\,6}^\infty$ of INSs as functions of $T_{\rm s}^\infty$ for our `most successful' models. Here, $T_{\rm s\,6}^\infty$ is the effective surface temperature expressed in MK, so that $dP/dT_{\rm s\,6}^\infty$ is the probability to find an INS with $T_{\rm s}^\infty$ lying in 1 MK interval. Because statistics of the observed sources is poor, we have averaged $\mathrm{d}P/\mathrm{d}T_{\rm s\,6}^\infty$ over the age interval from $t=10^{3.5}$ to
$10^{4.5}$ yr (as an example). The dashed line is for the HHJ EOS and the mass distribution 1i ($\alpha=0.1$, as in Fig.\ \ref{fig:CoolD_10_HHJ}) while the solid line is for the BSk21 EOS and the mass distribution 2i ($\alpha=0.09$, as in Fig.\ \ref{fig:CoolD_11_BSk21}). 

Even in the given, rather wide age interval there are only six observed sources. They are the neutron star in Pup A (source 2), PSR J1357--6429 (3), the Vela pulsar (4), PSR B1706--44 (5), PSR J0538+2817 (6), and XMMU J1732--3445 (15).  As discussed above, sources 2 and 15 can be explained by the cooling theory assuming strong proton superfluidity in the neutron star core
(e.g., \citealt{Klochkov_etal15}; Paper I) which we do not include in our current cooling models, for simplicity. Therefore, our present theory is relevant only for four sources, 3--6, whose positions are shown in Fig.\ \ref{fig:Probab} by thin vertical dotted lines. With this, extremely poor statistics, we cannot reliably estimate the differential probability distribution of the observed sources to be directly compared with theoretical ones. One can attempt to perform such an analysis using, for instance, Bayesian statistics (e.g., \citealt{Bayesian13}) which is beyond the scope of this work. Nevertheless, according to Fig.\ \ref{fig:Probab} the obtained theoretical probabilities seem in reasonable accord with the data.

We have plotted a number of other calculated probability distributions of isolated and accreting neutron stars (for different ages $t$ and mass accretion rates $\langle \dot{M} \rangle$, for both EOSs, different $M_{\rm D}$, $\alpha$ and mass distribution functions). We do not present them here. The possibility of accurate comparison with observations is as questionable as in the above example. The statistics of XRTs is better but the values $\langle \dot{M} \rangle$ and $L_\gamma^\infty$ extracted from observations are less certain. Our theoretical probability distributions seem to agree with the data but only qualitatively. Our present qualitative analysis `by eye' seems adequate at this stage of the investigation but more advanced analysis is required in future studies.

\begin{figure*}
\centering
\includegraphics[width=0.45\textwidth]{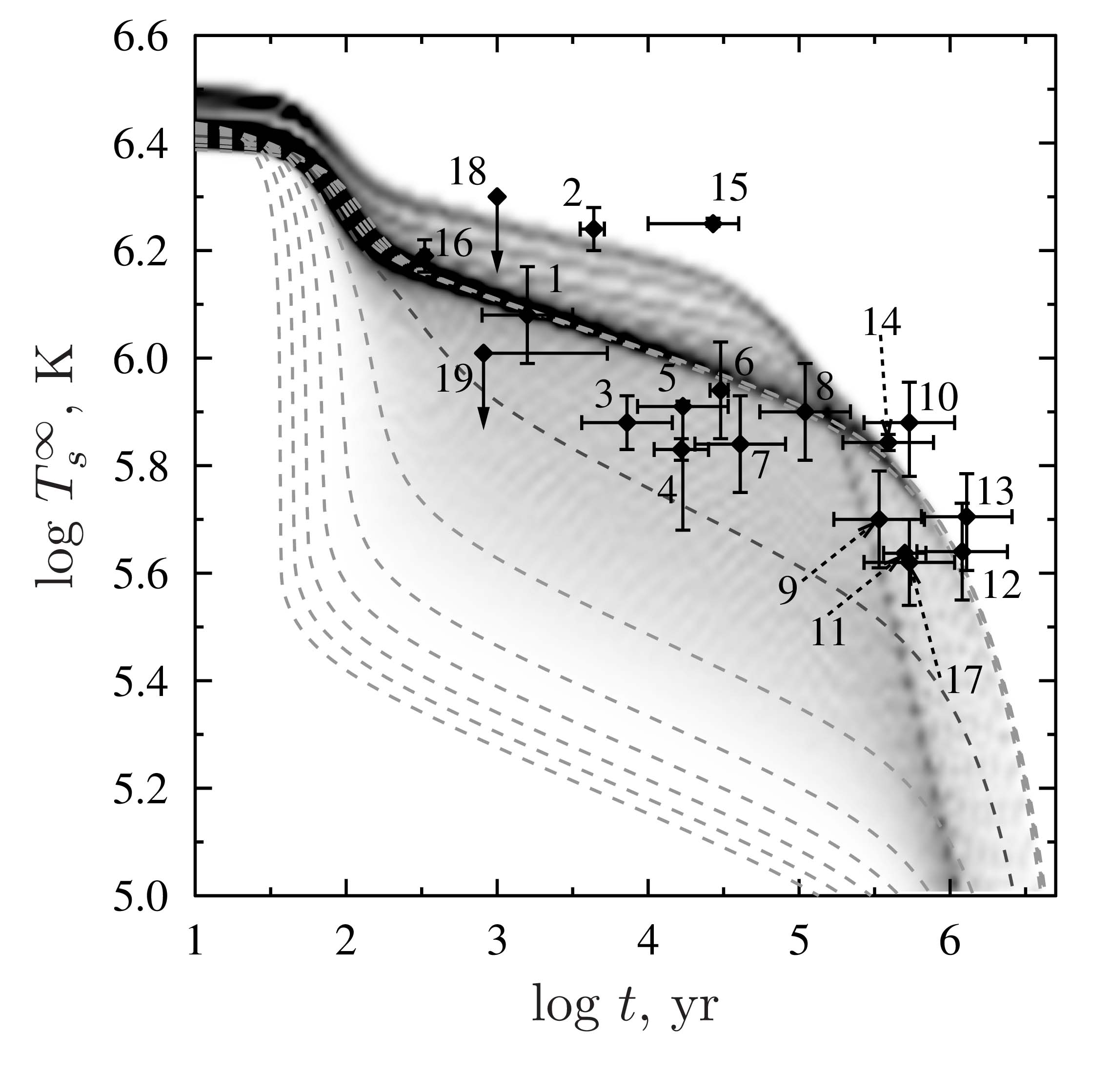}%
\includegraphics[width=0.45\textwidth]{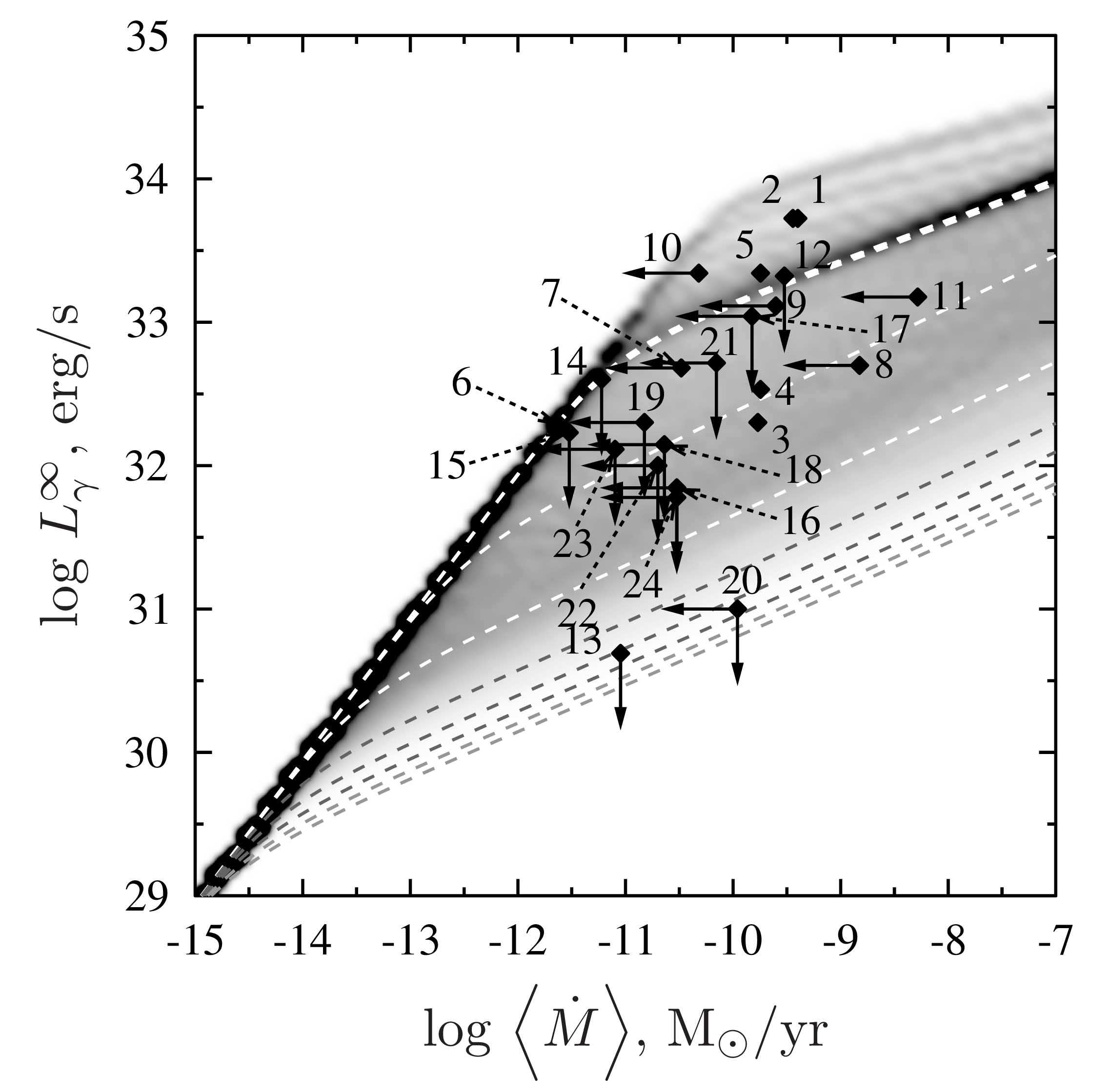}
\caption{Same as in Figs.\ \ref{fig:CoolD_10_HHJ} and \ref{fig:CoolD_11_BSk21} but for the BSk21 EOS, $\alpha=0.05$, $M_\mathrm{D}=1.65$ M$\odot$, and mass distributions 3i and 3a.}
\label{fig:CoolD_20_2_BSk21}
\end{figure*}
\begin{figure*}
\centering
\includegraphics[width=0.45\textwidth]{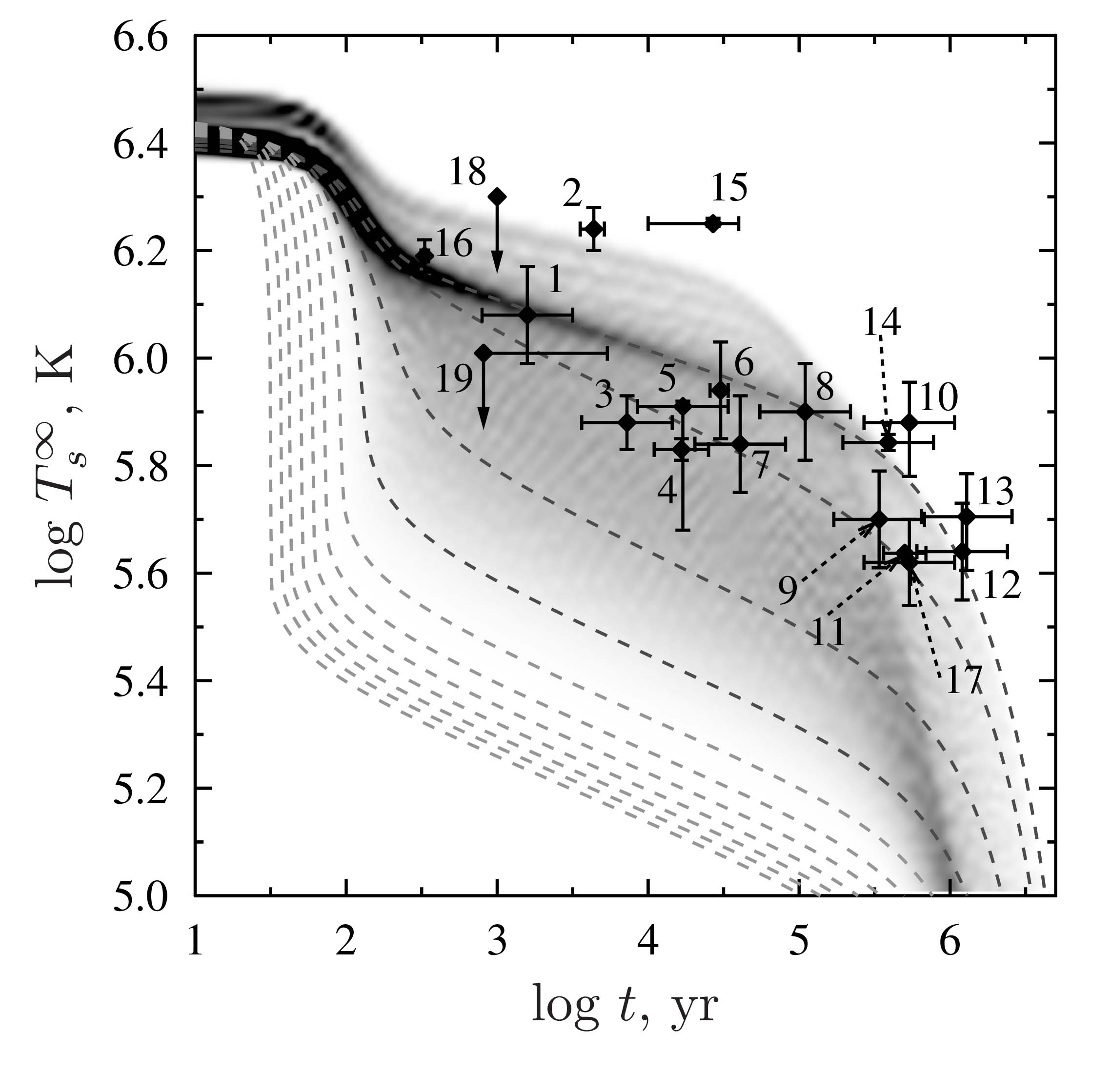}%
\includegraphics[width=0.45\textwidth]{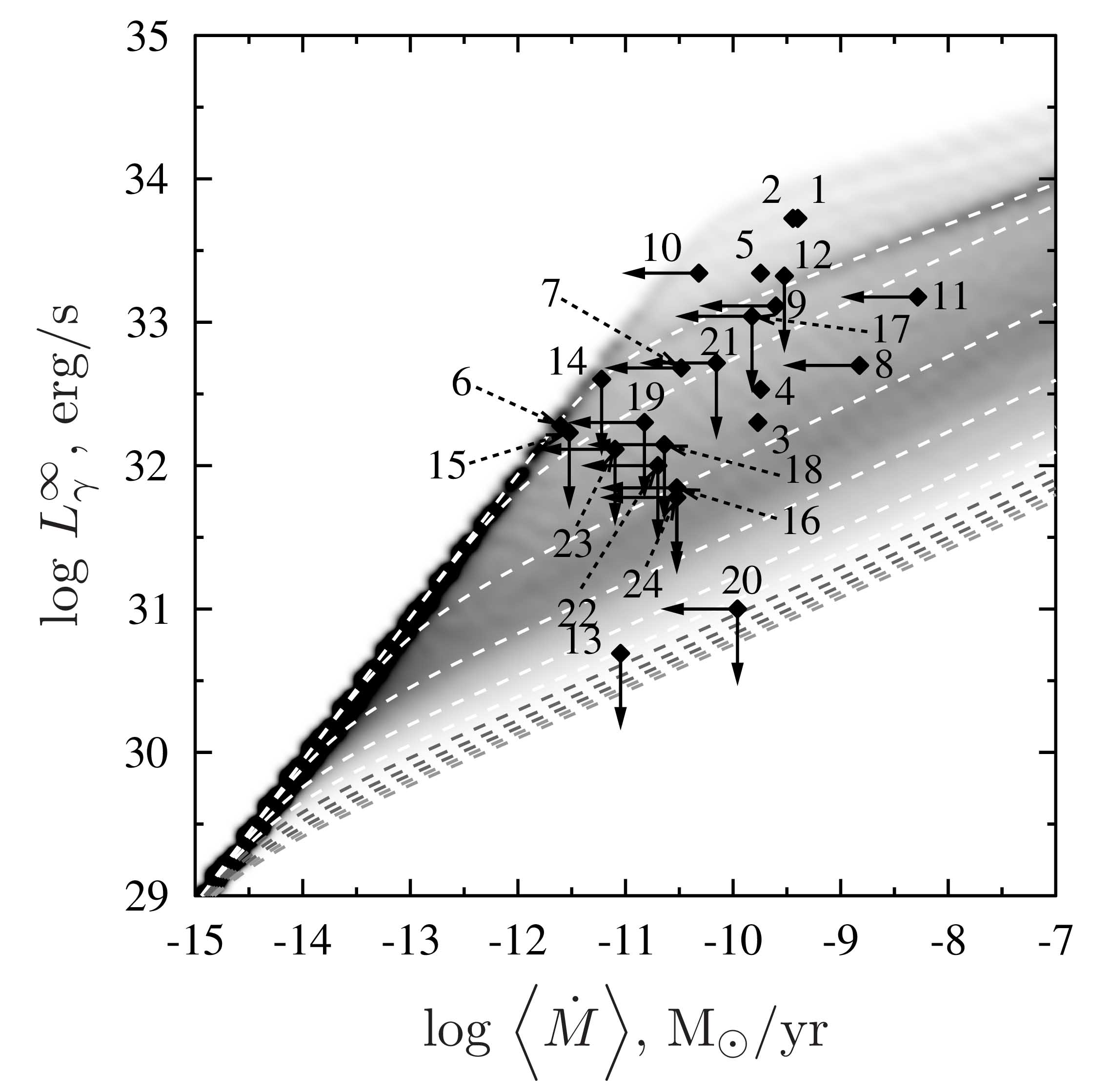}
\caption{Same as in Figs.\ \ref{fig:CoolD_10_HHJ}, \ref{fig:CoolD_11_BSk21} and \ref{fig:CoolD_20_2_BSk21} but for the BSk21 EOS, $\alpha=0.05$, $M_\mathrm{D}=1.45$ M$\odot$, and mass distributions 4i and 4a.} \label{fig:CoolD_20_1_BSk21}
\end{figure*}

\subsection{Less `successful' smaller broadening, $0.05 \lesssim \alpha \lesssim 0.08$}
\label{sec:mediumalpha}

Very small broadening of the direct Urca threshold ($\alpha$ smaller than 0.05) is not acceptable in our model (Section \ref{sec:smallalpha}) while the case of $\alpha \approx 0.08--0.10$ seems most suitable (Section \ref{sec:bestalpha}). Let us discuss the intermediate case $0.05 \lesssim \alpha \lesssim 0.08$. In this case, the theory can explain the observations but explanations become
less probable with decreasing $\alpha$ from $\alpha \approx 0.08$ to $\alpha \approx 0.05$ as far as mass distributions are concerned as discussed below. On the other hand, when $\alpha$ decreases the range of allowable $M_{\rm D}$ widens (Section \ref{sec:DurcaMass}).

We illustrate these statements in Figs.\ \ref{fig:CoolD_20_2_BSk21} and \ref{fig:CoolD_20_1_BSk21} by two examples for the lowest $\alpha \approx 0.05$. Fig.\ \ref{fig:CoolD_20_2_BSk21} presents the probability distributions for INSs and XRTs assuming the BSk21 EOS, $\alpha=0.05$, $M_\mathrm{D}=1.65$ M$\odot$, and the mass functions
3i and 3a (cf. with Figs.\ \ref{fig:CoolD_10_HHJ} and \ref{fig:CoolD_11_BSk21}). The critical mass for opening the direct Urca process is almost the same as in our best model for the BSk21
EOS (Fig.\ \ref{fig:CoolD_11_BSk21}). However, to explain all the data we now have to take narrower mass distributions 3i and 3a (Fig.\ \ref{fig:mfun}) which seem less probable than the mass
distributions 2i and 2a for the best model. Note also the non-uniformity of the `reference' curves, especially if compared with `most successful' BSk21 model (Fig.\ \ref{fig:CoolD_11_BSk21}).

Fig.\ \ref{fig:CoolD_20_2_BSk21} shows the probability distributions for INSs and XRTs assuming the BSk21 EOS, $\alpha=0.05$, and $M_\mathrm{D}=1.45$ M$\odot$. Here, we have taken the lowest $M_\mathrm{D}$ allowed at $\alpha=0.05$ (Fig.\ \ref{fig:BSk21-D-Mass}). Again, to explain the data we should take rather narrow mass distributions 4i and 4a (Fig.\ \ref{fig:mfun}) shifted to lower masses with respect to the previous mass distributions. This model, although formally possible, seems rather unlikely, because of three reasons. First, the required mass
distributions are too narrow; secondly, characteristic neutron star masses become uncomfortably low; thirdly, $M_{\rm D}$ is much lower than the mass $M_{\rm D0}$ determined by the BSk21 EOS. Large differences of $M_{\rm D}$ from $M_{\rm D0}$ would be difficult to explain from theoretical point of view.

Therefore, when we decrease $\alpha$ below $\alpha\approx 0.1$ but keep $M_{\rm D}\approx M_{\rm D0}$, we have to narrow neutron star mass distributions (which does not seem likely). If we, additionally, take lower $M_{\rm D}$ we increase the difference between $M_{\rm D}$ and $M_{\rm D0}$ and reduce typical masses of neutron stars (making our models even less likely).

Note that the majority of INSs should have masses $M \lesssim M_{\rm D}$. On the other hand, some neutron stars in XRTs should be more massive than $M_{\rm D}$ (Fig.\ \ref{fig:mfun}). First of all, this is needed to explain the observations of SAX J1804.4--3658  (source 13; \citealt{Campana_etal05,GC06,Heinke09b}) which indicate the operation of direct Urca process.

\subsection{Wider broadening, $\alpha \gtrsim 0.12$}
\label{sec:largealpha}

For completeness of our consideration let us discuss the case of relatively large broadening of the direct Urca threshold, $\alpha \gtrsim 0.12$. The general trend is clear. If we broaden the threshold, the direct Urca process operates in wider regions of neutron star cores and makes the stars colder. This destroys the agreement of the theory with observations. Formally, it is reflected in
the fact (Figs.\ \ref{fig:HHJ-D-Mass}-- \ref{fig:HHJ-BSk21-D-Mass}) that at $\alpha \gtrsim 0.12$ we obtain $M_{\rm D\, min}>M_{\rm D\,max}$. In other words, our qualitative analysis of Section \ref{sec:DurcaMass} signals that successful solutions of the problem disappear. For instance, as seen from figs 15 and 16 of Paper I plotted for the HHJ EOS at $M_{\rm D}=M_{\rm D0}$ and $\alpha=0.2$, the theory cannot explain the data.

Finally, we mention that at $M_{\mathrm{D\,max}}\! < \! M_{\mathrm{D\,min}}$ three cases are formally possible, (i) $M_{\mathrm{D}}\! < \! M_{\mathrm{D\,max}}\! < \! M_{\mathrm{D\,min}}$; (ii) $M_{\mathrm{D\,max}}\! < \! M_{\mathrm{D\,min}}\! < \! M_{\mathrm{D}}$ and (iii) $M_{\mathrm{D\,max}}\! < \! M_{\mathrm{D}}\! < \! M_{\mathrm{D\,min}}$. Case (i) is similar to what is shown in Paper I; theoretical domains of `warm' stars (in successful explanations) become cooler but domains of `cold' stars remain unchanged. Case (ii) is the opposite; `warmer' domains remain unchanged but `colder' ones become slightly warmer. The last case (iii) is a combination of two previous ones; domains of `warmer' stars become cooler and `colder' stars warmer. It is unlikely to reach satisfactory agreement with the observations of INSs and XRTs in all these cases. It would also be difficult to explain relatively large broadening of the direct Urca threshold from theoretical point of view.

\section{Conclusions}
\label{sec:concl}

We have extended the statistical approach of Paper I to study thermal evolution of middle-aged INSs and old accreting quasistationary neutron stars in XRTs in two respects.

First, we have considered two EOSs in neutron star cores (the HHJ and BSk21 EOSs, Table \ref{tab:models}) instead of one (HHJ) EOS in Paper I. Secondly, we have allowed the theoretical density threshold $\rho_{\rm D}$ for the direct Urca onset to be not only broadened (as described by the parameter $\alpha\approx \Delta \rho_{\rm D}/\rho_{\rm D0}$) but also shifted from its position $\rho_{\rm D0}$ determined by a given EOS. In this way we have actually extended a class of studied EOSs because in many cases theoretical EOSs give almost the same pressure $P(\rho)$ and neutron star models $M(R)$ but have sufficiently different direct Urca thresholds $\rho_{\rm D}$ (compare, for instance, the BSk19, BSk20, and BSk21 EOSs; \citealt{GCP10, PCGD12,Potekhin_etal13}) .

We have confirmed the principal conclusion of Paper I that the theory can explain observations of INSs and XRTs assuming the onset of direct Urca process in sufficiently massive stars with the broadened density threshold.

The main conclusions of the present investigation are as follows.

\begin{enumerate}

\item
At too small ($\alpha \lesssim 0.05$) and too high ($\alpha \gtrsim 0.12$) broadening a successful explanation of the data is unlikely.

\item
At intermediate broadening ($0.05 \lesssim \alpha \lesssim 0.12$) one can explain the data in a certain interval of critical masses $M_{\rm D}$ ($M_{\rm D\,min} \lesssim M_{\rm D} \lesssim M_{\rm D\,max}$) of stars, where the direct Urca sets in, by tuning the mass distributions $f_{\rm i}(M)$  and $f_{\rm a}(M)$ of isolated and accreting neutron stars. The results do not depend significantly on EOS (HHJ or BSk21). The majority of INSs should have masses $M \lesssim M_{\rm D}$, but some accreting neutron stars should be more massive.

\item
The maximum mass  $M_{\rm D\,max}\approx 1.7-1.8\,{\rm M}\odot$ is almost independent of $\alpha$, while the minimum mass $M_{\rm D\,min}$ decreases linearly with decreasing $\alpha$.

\item
Most `successful' explanations correspond to $\alpha \approx 0.08-0.10$ and $M_{\rm D} \approx 1.6-1.8\,{\rm M}\odot$, in which case $M_{\rm D\,min}$ is close to  $M_{\rm D\,max}$ and the mass distributions  $f_{\rm i}(M)$  and $f_{\rm a}(M)$ 
resemble distributions 1 and 2 in Fig.\ \ref{fig:mfun}.

\item
When $\alpha$ decreases from `most successful' values $\alpha \sim 0.1$ to the lowest values $\alpha \approx 0.05$, the minimum mass $M_{\rm D\,min}$ decreases and allowable interval of $M_{\rm D}$ becomes wider whereas the mass distributions $f_{\rm i}(M)$  and $f_{\rm a}(M)$ become uncomfortably narrow (like distributions 3 and 4 in Fig.\ \ref{fig:mfun}). If, in addition, $M_{\rm D}$ is taken much below $\approx 1.6\,{\rm M}\odot$, one gets uncomfortably low
neutron star masses.

\item
Large shifts of $M_{\rm D}$ from the value $M_{\rm D0}$, determined by a given EOS, and large broadenings $\alpha$ of the direct Urca threshold, as well as too narrow mass distributions   
$f_{\rm i}(M)$ and $f_{\rm a}(M)$  are unlikely from theoretical point of view; such formal solutions have to be regarded as less probable. With this in mind, 
the values $\alpha \approx 0.08-0.10$ and $M_{\rm D}\approx 1.6--1.8\,{\rm M}\odot$ seem
more probable than the others.

\end{enumerate}

Let us stress that our limitations of $M_{\rm D}$ are not rigorous (rather a number of indirect evidences). Our conclusions have to be regarded as preliminary and  can be refined. First of all, it would be instructive to replace our phenomenological model of shifted and broadened direct Urca threshold by some physical models taking into account the effects of superfluidity and/or magnetic fields as well as possible effects of nuclear physics (e.g., \citealt{BY99,YKGH01}, and references therein). In addition, one can try different mass distributions of cooling and accreting neutron stars (e.g., \citealt{Posselt_etal08,Kiziltan_etal13}) and formalize the comparison of theoretical and observational probability distributions of isolated and accreting neutron stars using accurate statistical methods (e.g., \citealt{Bayesian13}) instead of comparison `by eye'. It would be good to implement observational selection effects in this analysis (which seems to be a very complicated problem). These refinements would hopefully clarify the properties of the direct Urca process in neutron stars which are closely related to fundamental properties of superdense matter in neutron star cores. However, such improvements are beyond the scope of this paper. 
Note that statistical studies of evolution of neutron stars have been undertaken previously (e.g., \citealt{PGTB06,Posselt_etal08}) but using quite different approaches.

\section*{acknowledgements}
The authors are grateful to S.A. Balashev, J. N\"attil\"a and J. Poutanen 
for useful comments. The work of MB was partly supported by the Dynasty Foundation and Foundation for Support of Education and Science (Alferov's Foundation). The work of DY was
partly supported by Russian Foundation for Basic Research (grants Nos. 14-02-00868-a and 13-02-12017-ofi-M) and by `NewCompStar', COST Action MP1304.

%\bibliographystyle{mn2e}
%\bibliography{MyBibList}

\begin{thebibliography}{32}
\expandafter\ifx\csname natexlab\endcsname\relax\def\natexlab#1{#1}\fi

\bibitem[{{Baiko} \& {Yakovlev}(1999)}]{BY99}
{Baiko} D.~A., {Yakovlev} D.~G., 1999, Astron. Astrophys., 342, 192

\bibitem[{{Beznogov} \& {Yakovlev}(2015)}]{BY15}
{Beznogov} M.~V., {Yakovlev} D.~G., 2015, MNRAS, 447, 1598

\bibitem[{{Blaschke}, {Grigorian} \& {Voskresensky}(2004){Blaschke},
  {Grigorian}, \& {Voskresensky}}]{Blaschke_etal04}
{Blaschke} D., {Grigorian} H., {Voskresensky} D.~N., 2004, Astron. Astrophys.,
  424, 979

\bibitem[{{Brown}, {Bildsten} \& {Rutledge}(1998){Brown}, {Bildsten}, \&
  {Rutledge}}]{BBR98}
{Brown} E.~F., {Bildsten} L., {Rutledge} R.~E., 1998, Astrophys. J. Lett., 504,
  L95

\bibitem[{{Campana} {et~al}\mbox{.}(2002){Campana}, {Stella}, {Gastaldello},
  {Mereghetti}, {Colpi}, {Israel}, {Burderi}, {Di Salvo}, \&
  {Robba}}]{Campana_etal05}
{Campana} S. {et~al.}, 2002, Astrophys. J. Lett., 575, L15

\bibitem[{{Galloway} \& {Cumming}(2006)}]{GC06}
{Galloway} D.~K., {Cumming} A., 2006, Astrophys. J., 652, 559

\bibitem[{{Gelman} {et~al}\mbox{.}(2013){Gelman}, {Carlin}, {Stern}, {Dunson},
  {Vehtari}, \& {Rubin}}]{Bayesian13}
{Gelman} A., {Carlin} J.~B., {Stern} H.~S., {Dunson} D.~B., {Vehtari} A.,
  {Rubin} D.~B., 2013, {Bayesian Data Analysis}. CRC Press, New York

\bibitem[{{Goriely}, {Chamel} \& {Pearson}(2010){Goriely}, {Chamel}, \&
  {Pearson}}]{GCP10}
{Goriely} S., {Chamel} N., {Pearson} J.~M., 2010, Phys. Rev. C, 82, 035804

\bibitem[{{Gudmundsson}, {Pethick} \& {Epstein}(1983){Gudmundsson}, {Pethick},
  \& {Epstein}}]{GPE83}
{Gudmundsson} E.~H., {Pethick} C.~J., {Epstein} R.~I., 1983, Astrophys. J.,
  272, 286

\bibitem[{{Haensel}, {Potekhin} \& {Yakovlev}(2007){Haensel}, {Potekhin}, \&
  {Yakovlev}}]{HPY07}
{Haensel} P., {Potekhin} A.~Y., {Yakovlev} D.~G., 2007, Astrophysics and Space
  Science Library, Vol. 326, {Neutron Stars. 1. Equation of State and
  Structure}. Springer, New York

\bibitem[{{Haensel} \& {Zdunik}(1990)}]{HZ90}
{Haensel} P., {Zdunik} J.~L., 1990, Astron. Astrophys., 227, 431

\bibitem[{{Haensel} \& {Zdunik}(2008)}]{HZ08}
{Haensel} P., {Zdunik} J.~L., 2008, Astron. Astrophys., 480, 459

\bibitem[{{Heinke} {et~al}\mbox{.}(2009){Heinke}, {Jonker}, {Wijnands},
  {Deloye}, \& {Taam}}]{Heinke09b}
{Heinke} C.~O., {Jonker} P.~G., {Wijnands} R., {Deloye} C.~J., {Taam} R.~E.,
  2009, Astrophys. J., 691, 1035

\bibitem[{{Heiselberg} \& {Hjorth-Jensen}(1999)}]{HHJ99}
{Heiselberg} H., {Hjorth-Jensen} M., 1999, Astrophys. J. Lett., 525, L45

\bibitem[{{Kaminker} {et~al}\mbox{.}(2014){Kaminker}, {Kaurov}, {Potekhin}, \&
  {Yakovlev}}]{KKPY14}
{Kaminker} A.~D., {Kaurov} A.~A., {Potekhin} A.~Y., {Yakovlev} D.~G., 2014,
  MNRAS, 442, 3484

\bibitem[{{Kiziltan} {et~al}\mbox{.}(2013){Kiziltan}, {Kottas}, {De Yoreo}, \&
  {Thorsett}}]{Kiziltan_etal13}
{Kiziltan} B., {Kottas} A., {De Yoreo} M., {Thorsett} S.~E., 2013, Astrophys.
  J., 778, 66

\bibitem[{{Kl{\"a}hn} {et~al}\mbox{.}(2006){Kl{\"a}hn}, {Blaschke}, {Typel},
  {van Dalen}, {Faessler}, {Fuchs}, {Gaitanos}, {Grigorian}, {Ho},
  {Kolomeitsev}, {Miller}, {R{\"o}pke}, {Tr{\"u}mper}, {Voskresensky}, {Weber},
  \& {Wolter}}]{Blaschke_etal06}
{Kl{\"a}hn} T. {et~al.}, 2006, Phys. Rev. C, 74, 035802

\bibitem[{{Klochkov} {et~al}\mbox{.}(2013){Klochkov}, {P{\"u}hlhofer},
  {Suleimanov}, {Simon}, {Werner}, \& {Santangelo}}]{Klochkov_etal13}
{Klochkov} D., {P{\"u}hlhofer} G., {Suleimanov} V., {Simon} S., {Werner} K.,
  {Santangelo} A., 2013, Astron. Astrophys., 556, A41

\bibitem[{{Klochkov} {et~al}\mbox{.}(2015){Klochkov}, {Suleimanov},
  {P{\"u}hlhofer}, {Yakovlev}, {Santangelo}, \& {Werner}}]{Klochkov_etal15}
{Klochkov} D., {Suleimanov} V., {P{\"u}hlhofer} G., {Yakovlev} D.~G.,
  {Santangelo} A., {Werner} K., 2015, Astron. Astrophys., 573, A53

\bibitem[{{Lattimer} {et~al}\mbox{.}(1991){Lattimer}, {Pethick}, {Prakash}, \&
  {Haensel}}]{LPPH91}
{Lattimer} J.~M., {Pethick} C.~J., {Prakash} M., {Haensel} P., 1991, Phys. Rev.
  Lett., 66, 2701

\bibitem[{{Page} {et~al}\mbox{.}(2004){Page}, {Lattimer}, {Prakash}, \&
  {Steiner}}]{PLPS04}
{Page} D., {Lattimer} J.~M., {Prakash} M., {Steiner} A.~W., 2004, Astrophys. J.
  Suppl., 155, 623

\bibitem[{{Page} {et~al}\mbox{.}(2009){Page}, {Lattimer}, {Prakash}, \&
  {Steiner}}]{Page_etal09}
{Page} D., {Lattimer} J.~M., {Prakash} M., {Steiner} A.~W., 2009, Astrophys.
  J., 707, 1131

\bibitem[{{Pearson} {et~al}\mbox{.}(2012){Pearson}, {Chamel}, {Goriely}, \&
  {Ducoin}}]{PCGD12}
{Pearson} J.~M., {Chamel} N., {Goriely} S., {Ducoin} C., 2012, Phys. Rev. C,
  85, 065803

\bibitem[{{Popov} {et~al}\mbox{.}(2006){Popov}, {Grigorian}, {Turolla}, \&
  {Blaschke}}]{PGTB06}
{Popov} S., {Grigorian} H., {Turolla} R., {Blaschke} D., 2006, Astron.
  Astrophys., 448, 327

\bibitem[{{Posselt} {et~al}\mbox{.}(2008){Posselt}, {Popov}, {Haberl},
  {Tr{\"u}mper}, {Turolla}, \& {Neuh{\"a}user}}]{Posselt_etal08}
{Posselt} B., {Popov} S.~B., {Haberl} F., {Tr{\"u}mper} J., {Turolla} R.,
  {Neuh{\"a}user} R., 2008, Astron. Astrophys., 482, 617

\bibitem[{{Potekhin}, {Chabrier} \& {Yakovlev}(1997){Potekhin}, {Chabrier}, \&
  {Yakovlev}}]{Potekhin_etal97}
{Potekhin} A.~Y., {Chabrier} G., {Yakovlev} D.~G., 1997, Astron. Astrophys.,
  323, 415

\bibitem[{{Potekhin} {et~al}\mbox{.}(2013){Potekhin}, {Fantina}, {Chamel},
  {Pearson}, \& {Goriely}}]{Potekhin_etal13}
{Potekhin} A.~Y., {Fantina} A.~F., {Chamel} N., {Pearson} J.~M., {Goriely} S.,
  2013, Astron. Astrophys., 560, A48

\bibitem[{{Schaab} {et~al}\mbox{.}(1997){Schaab}, {Voskresensky}, {Sedrakian},
  {Weber}, \& {Weigel}}]{Schaab_etal97}
{Schaab} C., {Voskresensky} D., {Sedrakian} A.~D., {Weber} F., {Weigel} M.~K.,
  1997, Astron. Astrophys., 321, 591

\bibitem[{{Yakovlev} \& {Haensel}(2003)}]{YH03}
{Yakovlev} D.~G., {Haensel} P., 2003, Astron. Astrophys., 407, 259

\bibitem[{{Yakovlev} {et~al}\mbox{.}(2001){Yakovlev}, {Kaminker}, {Gnedin}, \&
  {Haensel}}]{YKGH01}
{Yakovlev} D.~G., {Kaminker} A.~D., {Gnedin} O.~Y., {Haensel} P., 2001, Phys.
  Rep., 354, 1

\bibitem[{{Yakovlev}, {Levenfish} \& {Haensel}(2003){Yakovlev}, {Levenfish}, \&
  {Haensel}}]{YLH03}
{Yakovlev} D.~G., {Levenfish} K.~P., {Haensel} P., 2003, Astron. Astrophys.,
  407, 265

\bibitem[{{Yakovlev} \& {Pethick}(2004)}]{YP04}
{Yakovlev} D.~G., {Pethick} C.~J., 2004, Annu. Rev. Astron. Astrophys., 42, 169

\end{thebibliography}

\end{document}